# The Zwicky Transient Facility: Data Processing, Products, and Archive

Frank J. Masci[1], Russ R. Laher[1], Ben Rusholme[1], David L. Shupe[1], Steven Groom[1], Jason Surace[1], Edward Jackson[1], Serge Monkewitz[1], Ron Beck[1], David Flynn[1], Scott Terek[1], Walter Landry[1], Eugean Hacopians[1], Vandana Desai[1], Justin Howell[1], Tim Brooke[1], David Imel[1], Stefanie Wachter[2], Quan-Zhi Ye[1,3], Hsing-Wen Lin[4,5], S. Bradley Cenko[6], Virginia Cunningham[7], Umaa Rebbapragada[8], Brian Bue[8], Adam A. Miller[9,10], Ashish Mahabal[3], Eric C. Bellm[11], Maria T. Patterson[11], Mario Jurić[11], V. Zach Golkhou[11,12,18], Eran O. Ofek[13], Richard Walters[14], Matthew Graham[3], Mansi M. Kasliwal[3], Richard G. Dekany[14], Thomas Kupfer[3,15], Kevin Burdge[3], Christopher B. Cannella[3], Tom Barlow[3], Angela Van Sistine[16], Matteo Giomi[17], Christoffer Fremling[3], Nadejda Blagorodnova[3], David Levitan[3], Reed Riddle[14], Roger M. Smith[14], George Helou[1], Thomas A. Prince[3], and Shrinivas R. Kulkarni[3]

[1] IPAC, California Institute of Technology, 1200 E. California Boulevard, Pasadena, CA 91125, USA; fmasci@caltech.edu
[2] Carnegie Observatories, 813 Santa Barbara Street, Pasadena, CA 91101, USA
[3] Division of Physics, Mathematics and Astronomy, California Institute of Technology, Pasadena, CA 91125, USA
[4] Department of Physics, University of Michigan, Ann Arbor, MI 48109, USA
[5] Institute of Astronomy, National Central University, 32001, Taiwan
[6] Astrophysics Science Division, NASA/Goddard Space Flight Center, MC 661, Greenbelt, MD 20771, USA
[7] Department of Astronomy, University of Maryland, Stadium Drive, College Park, MD 20742, USA
[8] Jet Propulsion Laboratory, California Institute of Technology, Pasadena, CA 91109, USA
[9] Center for Interdisciplinary Exploration and Research in Astrophysics and Department of Physics and Astronomy, Northwestern University, 2145 Sheridan Road, Evanston, IL 60208, USA
[10] The Adler Planetarium, Chicago, IL 60605, USA
[11] DIRAC Institute, Department of Astronomy, University of Washington, Seattle, WA 98195, USA
[12] The eScience Institute, University of Washington, Seattle, WA 98195, USA
[13] Benoziyo Center for Astrophysics, Weizmann Institute of Science, 76100 Rehovot, Israel
[14] Caltech Optical Observatories, California Institute of Technology, Pasadena, CA 91125, USA
[15] KITP, Kohn Hall, University of California, Santa Barbara, CA 93106, USA
[16] Department of Physics, University of Wisconsin, Milwaukee, WI 53201, USA
[17] DESY, 15735 Zeuthen, Germany
Received 2018 July 14; accepted 2018 October 15; published 2018 December 7

## Abstract

The Zwicky Transient Facility (ZTF) is a new robotic time-domain survey currently in progress using the Palomar 48-inch Schmidt Telescope. ZTF uses a 47 square degree field with a 600 megapixel camera to scan the entire northern visible sky at rates of ∼3760 square degrees/hour to median depths of $g \sim 20.8$ and $r \sim 20.6$ mag (AB, $5\sigma$ in 30 sec). We describe the Science Data System that is housed at IPAC, Caltech. This comprises the data-processing pipelines, alert production system, data archive, and user interfaces for accessing and analyzing the products. The real-time pipeline employs a novel image-differencing algorithm, optimized for the detection of point-source transient events. These events are vetted for reliability using a machine-learned classifier and combined with contextual information to generate data-rich alert packets. The packets become available for distribution typically within 13 minutes (95th percentile) of observation. Detected events are also linked to generate candidate moving-object tracks using a novel algorithm. Objects that move fast enough to streak in the individual exposures are also extracted and vetted. We present some preliminary results of the calibration performance delivered by the real-time pipeline. The reconstructed astrometric accuracy per science image with respect to *Gaia* DR1 is typically 45 to 85 milliarcsec. This is the RMS per-axis on the sky for sources extracted with photometric S/N $\geqslant$ 10 and hence corresponds to the typical astrometric uncertainty down to this limit. The derived photometric precision (repeatability) at bright unsaturated fluxes varies between 8 and 25 millimag. The high end of these ranges corresponds to an airmass approaching ∼2—the limit of the public survey. Photometric calibration accuracy with respect to Pan-STARRS1 is generally better than 2%. The products support a broad range of scientific applications: fast and young supernovae; rare flux transients; variable stars; eclipsing binaries; variability from active galactic nuclei;

---

[18] Moore-Sloan, WRF, and DIRAC Fellow.

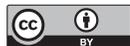

counterparts to gravitational wave sources; a more complete census of Type Ia supernovae; and solar-system objects.

*Online material:* color figures

## 1. Introduction

The Zwicky Transient Facility (ZTF)[19] is a next-generation optical time-domain survey currently in operation. It builds upon the predecessor surveys: the Palomar Transient Factory (PTF; Law et al. 2009; Rau et al. 2009) followed by the *intermediate* PTF (iPTF) program using the Palomar 48-inch Schmidt Telescope (P48) at Palomar Observatory. ZTF is extending our knowledge of the temporal and dynamic sky. This includes near-Earth asteroids (NEAs); rare and fast-evolving flux transients; and all classes of Galactic variable sources. An overview of the science goals is given in Graham et al. (2018). Technical specifications of the observing system (OS), its performance, survey design, and early science results are described in Bellm et al. (2019). More detailed descriptions of the OS instrumentation are given in Dekany et al. (2016) and Dekany et al. (2018).

This paper gives a high-level overview of the ZTF Science Data System (hereafter ZSDS) housed at IPAC[20], Caltech. Core functions of the ZSDS include managing data transfer from the P48; raw data ingestion; all processing pipelines; long-term archiving and curation of data products; user interfaces for data retrieval and access management; near-real-time distribution of flux-transient alerts and potentially new solar-system objects (SSOs); generation of quality assurance (QA) metrics for the project; analysis and trending; maintenance of all software, hardware, and databases; and user support. We review the delivered science products and services for accessing and analyzing those products. We also present some preliminary results of the astrometric and photometric accuracy achieved by the core processing pipelines.

Mahabal et al. (2018) describe the machine-learned (ML) vetting subsystems that interface with the ZSDS pipelines and their performance in purifying the raw transient events extracted from them. Pipeline algorithms, data product specifics, formats, and usage are described in more detail in the ZSDS Explanatory Supplement located at http://www.ztf.caltech.edu/page/technical#science-data-system, while access to the ZTF archive with example queries are described at https://irsa.ipac.caltech.edu/Missions/ztf.html. The first public data release is anticipated approximately one year after the start of survey operations[21], with further releases occurring every six months thereafter.

Section 2 reviews the ZSDS design: compute architecture, archive, databases, and operational tasks. The processing pipelines are described in Section 3, and all science data products derived therefrom are summarized in Section 4. Archive access methods, services, and analysis platforms are described in Section 5. Current data rates and volumes, source-extraction statistics, and expectations for the survey are summarized in Section 6. Performance of the real-time pipeline, both in terms of runtime; alert production statistics and latency; and quality of some of the archive products is discussed in Section 7. Some lessons learned and advisories are given in Section 8. Conclusions, work in progress, and additional planned functionality are summarized in Section 9. All acronyms are defined in Appendix A.

## 2. Data System and Operations Overview

The ZSDS was developed with one primary goal: to provide a processing and archival system that delivers science quality products as close to real time as possible. The requirement is 20 minutes from observation. This is apportioned into 10 minutes to transer the data to IPAC and another 10 minutes to process it. This was driven primarily to support the discovery of fast-evolving transients, and was set based on the rate of evolution of relevant transients, as well as the time required to trigger relevant follow-up observations. The ZSDS is a new development at IPAC, but capitalizes on the lessons learned from the PTF and iPTF data systems (Laher et al. 2014; Masci et al. 2017) to meet the $\simeq 15\times$ higher data rate from ZTF. Its exclusive in-house design and implementation made maintainability and adaptability to changing project needs tremendously flexible.

Research and development have proceeded in a highly cost-constrained[22] environment (Surace et al. 2015), and continue to be refined in response to feedback received from the partner science programs. Cost was the primary driver that determined the overall design of the ZSDS, in particular the various products available (Section 4), their formats, and the services for retrieving and analyzing them (Section 5).

The ZSDS heavily relies on a highly parallelized, high-throughput processing infrastructure that is integrated with distributed file servers and relational databases. The ZTF archive is hosted by the NASA/IPAC Infrared Science Archive (IRSA[23]). Laher et al. (2018) reviewed some of the ZSDS components during the development phase. Below, we summarize these components and their integration into the overall ZSDS.

---

[19] http://www.ztf.caltech.edu
[20] Formerly referred to as the Infrared Processing & Analysis Center: https://www.ipac.caltech.edu.
[21] The science survey officially commenced on 2018-03-17 (UT).

[22] The workforce was $\simeq 5.7$ full-time equivalents (FTEs) during the development phase (ending in 2017 September). This dropped to $\simeq 3.7$ FTEs during the commissioning and science-validation periods (ending in February 2018). The workforce during operations is 2.5 FTEs.
[23] https://irsa.ipac.caltech.edu

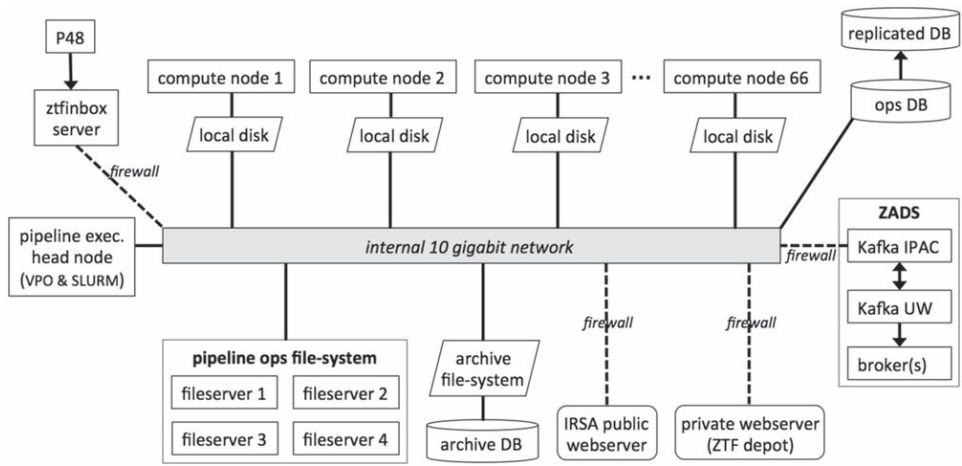

**Figure 1.** ZSDS infrastructure components and interfaces.

Figure 1 depicts the primary components of the ZSDS. It features 66 compute cluster nodes (see Section 2.1); four pipeline file servers connected to a file system; an archive file server connected to a separate file system; servers to support raw data ingestion, workload management, and monitoring; three database (DB) servers: the primary pipeline operations DB, the secondary (real-time-replicated) pipeline DB, and the archive DB; a private webserver referred to as ZTF-Depot (see below); a Kafka[24] cluster to support alert distribution (ZADS; Section 4.1); and the IRSA public webserver for distributing archived products. The local network can tolerate data rates of up to ten gigabits per second.

The real-time pipeline (Sections 3.5 and 3.6) generates many intermediate files. These interim files are written to the local node disks for speed and to avoid unnecessary congestion on the local network. The four pipeline file servers allow for efficient parallel file transfer across the local network during processing. Files copied from and to the pipeline file system generate the most traffic. Pipeline products are also copied to the archive file system in real time, i.e., immediately following the processing of a single-image unit (see below). This parallelized file server–compute architecture is also scalable. More servers can be added to accommodate increases in data rate, product volumes, and/or computing needs. The pipeline and archive file servers are spinning disk using the Network File System (NFS) protocol. This cost effective off-the-shelf solution achieves acceptable performance. We have not yet seen reasons to warrant a more sophisticated implementation.

### 2.1. Processing Architecture

The ZTF camera consists of 16 CCDs, each with $\sim 6000^2$ pixels. Furthermore, each CCD consists of four $\sim 3000^2$ pixel amplifier channels with overscan (bias) strips. We shall refer to these amplifiers as CCD-quadrants from hereon. Therefore, each camera exposure consists of 64 CCD-quadrants. The raw data files received at IPAC are CCD-based, where each file contains the already-split CCD-quadrants and corresponding overscans, each packaged into separate extensions of a compressed multi-extension FITS file (MEF; Section 3.3).

A CCD quadrant defines the basic image unit for concurrent (parallel) processing in the ZSDS and from which all data products are derived. This includes all calibrations, processed science images, difference images, reference (co-added) images, source catalogs, and light-curve files. All file products are CCD-quadrant based. There are no cross-quadrant or cross-CCD-dependent processing steps. All quadrants are instrumentally calibrated and treated independently end-to-end.

This CCD-quadrant-based processing approach is primarily driven by our experience with PTF. The ZTF CCD quadrants are of approximately the same size (in number of pixels and areal coverage) as the PTF CCDs. First, as was the case for PTF, most ZTF data products are distributed as flat data files (Section 4). Images this size generate file products (which include source catalog tables) that are easier to manage, analyze, and distribute using existing software and technologies. Second, there are processing steps that are determined to be more robust and accurate (in terms of reducing systematics from spatial variations) if performed as local as possible (spatially) on the P48's $7° \times 7°$ focal plane. For example, astrometric calibration, PSF derivation, photometric-zero-point derivation (Section 3.5), and PSF-matching prior to image differencing (Section 3.6). The CCD-quadrants are large enough to admit sufficient numbers of sources to optimize these steps (in the signal-to-noise (S/N) sense), but small enough to keep memory use, file product sizes, and hence processing throughput at a manageable level all over the sky during processing.

---
[24] https://kafka.apache.org

Table 1
Compute Cluster Specifications

| Quantity | Chip | Base Clock | vCPUs | RAM | Model |
| --- | --- | --- | --- | --- | --- |
| 32 | E5-2640v3 | 2.6 GHz | 32 | 128 GB | Dell R430 (2015) |
| 32 | E5-2640v4 | 2.4 GHz | 40 | 128 GB | Dell R430 (2017) |
| 2 | E5-2640v4 | 2.4 GHz | 40 | 768 GB | Supermicro |

Following splitting of the incoming raw data MEF files into CCD-quadrant images (Section 3.3), they are processed on a 66-node compute cluster. The cluster is composed of commodity 1U dual Xeon servers; 64 outfitted with 128 GB RAM and two high-memory nodes with 768 GB each. The latter are intended for executing ZTF's Moving Object Discovery Engine (ZMODE pipeline; Section 3.9), which performs memory intensive moving-object detection and track generation. The specifications of the compute cluster are summarized in Table 1. We have 1192 ($16 \times 32 + 34 \times 20$) *physical* processor cores across all compute nodes.

Due to data IO load limitations on the databases and archive file servers, we found that eight pipeline instances per node (or concurrent threads, where one thread represents the processing of one CCD quadrant) is optimal and sufficient for processing the incoming data rate in real time, with no latency in the delivery of products (see Section 7.1). By "optimal", we mean maximizing the product generation rate. Therefore, we have in total 528 ($8 \times 66$) concurrent threads processing the incoming data, which arrives at a rate of one exposure/40 sec or equivalently, 64 quadrants/40 sec or 96 quadrants/minute on average. A single quadrant image can be processed in under five minutes on average in terms of wall-clock time (Section 7.1). Therefore, a naïve estimate for the number of processor cores needed to match the incoming data rate is $\simeq inp\_rate / proc\_rate = 96/(1/5) = 480$. This is really a lower limit, as the processing time is not purely CPU, i.e., disk and network IO are not negligible. Given our good network bandwidth and disk speeds however, we have more than enough cores to handle the incoming data rate.

It is important to note that we are currently underutilizing our compute resources and hence have the capacity to do more. We have a total of 1192 physical cores, but we are currently using less than half. Additional nodes were purchased early in the survey to support ad hoc processing tasks, such as reference-image generation and moving-object detection. These tasks run asynchronously and not during real-time (nightly operations). Requirements on real-time processing and product delivery timescales are being met; however, there is a desire to increase the overall throughput. We are in the process of increasing the number of processor cores for real-time processing. This exercise entails exploring the additional load on the database and archive ingestion servers to ensure the higher production rates can be sustained throughout a night.

The compute cluster runs *CentOS7* and is managed using the SLURM[25] resource manager (packaged by the OPENHPC[26] project). The cluster runs a diverse set of workloads, and is configured with a number of partitions having different polices to maximize throughput. The nightly real-time processing runs in one of the highest-priority partitions, before almost all other pending work.

The Kafka (see footnote 6) cluster consists of an additional set of three compute nodes, each configured with ≈3 TB of local storage. This cluster facilitates the distribution of transient-alert packets to other institutions and eventually the community (the ZADS framework; Section 4.1). Currently, the IPAC Kafka cluster is mirrored close to real time with Kafka running at the University of Washington (UW) over the Internet.

ZTF-Depot is a private webserver to facilitate access to pipeline QA metrics, survey performance statistics, and survey depth-of-coverage maps (both cumulative and nightly, and movies thereof). It also serves as a portal to support the retrieval of products related to solar-system science and NEA discovery in near-real-time: primarily the outputs from streak detection (Section 3.6) and ZMODE (Section 3.9). All other pipeline products destined for public release are copied to the long-term archive at IRSA. These will be accessible through a separate webserver using either a graphical user interface (GUI) or application programming interface (API; Section 5).

### 2.2. Databases

The ZSDS pipelines are heavily reliant on a relational database for querying and storing the state (usability) of all intermediate and final file products. This includes QA metrics and observing system diagnostics for long-term trending and analysis. All DB servers in the ZSDS run PostgreSQL 9.6[27], and the primary pipeline operations DB server is equipped with 384 GB of memory.

Because the operations DB is not easily scalable in terms of IO or computing load, careful design and tuning of both the hardware and software were necessary. Database tables and indexes are arranged so that heavily accessed data are stored on solid-state devices (SSDs) as opposed to spinning disks. The higher memory of the operations DB server ensures a relatively high cache-hit rate. The largest DB table stores the raw transient candidates. This table is partitioned into child tables, with each pointing to a specific observing date and CCD quadrant. A set-up process that creates DB views over the correct candidate partitions is executed daily.

---
[25] https://slurm.schedmd.com
[26] http://openhpc.community
[27] https://www.postgresql.org

Database snapshots are copied to a network-attached file system, and snapshots covering four weeks of recent content are kept and available at any given time. Tape backups of the snapshotted data are separately made on a regular schedule, and moved to an off-site repository. The primary operations DB is also continuously replicated into a separate secondary DB for offline use. This secondary DB allows for analyses to be performed without impacting the performance of the primary DB during operations.

A separate relational DB keeps track of distilled archive product metadata on the IRSA (archive) side. This DB is used for user queries to search for specific archived file products (Sections 4 and 5.1). Another relational DB on the IRSA side is used to store source photometry and metadata extracted from the reference image co-adds (Section 3.7) with additional statistics computed by the source-matching pipeline (Section 3.8). This is referred to as the ZTF Objects Database and is queried to support the retrieval of light curves (multi-epoch photometry) using a file-based look-up (Section 5.3). There is no separate DB table that stores the contents of the epochal source catalog files (Section 4). This is because such a table will approach of order a trillion rows at the end of the nominal three-year survey. This was determined to be too costly to develop and maintain given the allocated resources.

### 2.3. Operational Routine and Tasks

Most of nightly processing and end-of-night tasks are orchestrated by an automated "watchdog" script referred to as the Virtual Pipeline Operator (VPO). This script checks for incoming raw data, creates all date-dependent paths in the operations, and ZTF-Depot filesystems, executes the Ingest pipeline (Section 3.2), and then the relevant calibration-generation or real-time science pipelines in accordance with the priorities programmed into the VPO. The VPO also accumulates QA statistics and number of transient-alerts extracted at regular intervals throughout the night and posts them to ZTF-Depot.

At the end of a night, the VPO performs accountability checks on the number of raw data files sent from the P48, received, successfully ingested (with all required metadata), split, processed, and archived. All statistics and metadata are accumulated in an end-of-night report and emailed to members of the project. Following nightly processing, the VPO triggers the ZMODE (moving-object) pipeline (Section 3.9); checks which fields and CCD quadrants have enough good-quality exposures to warrant reference image creation (if a reference image does not already exist) and, if so, triggers the reference-image pipeline

(Section 3.7) and updates the survey all-sky depth-of-coverage maps. The VPO then prepares the processing system for the forthcoming night by copying and partitioning the contents of the transient-candidates DB table into files to facilitate faster source matching with historical detections in real time when generating alert packets.

Other automated daily maintenance tasks involve monitoring disk space, removing old intermediate products from the operations file system, and database "vacuuming." Manual tasks involve monitoring all ZSDS hardware, archive access loads, throughput performance, tending to ad hoc reprocessing requests, and communicating any fatal system errors or outages to the project. Longer-term manual tasks involve executing the light-curve (source-matching) pipeline (Section 3.8) on typically monthly timescales, and performing synopses and quality checks of the reference-image holdings.

## 3. Pipelines

Figure 2 shows the core pipelines, their interdependencies, and how they interface with the archive, ZTF-Depot, and ZADS. A high-level summary of how these components are integrated into the overall architecture was given in Section 2. This section reviews the processing pipelines, but still only at a high-level. More detailed descriptions of the individual processing steps and algorithms are provided in the ZSDS Explanatory Supplement (referenced in Section 1).

There are nine pipelines in total. Seven are depicted in Figure 2, as raw data ingestion and splitting are actually two separate pipelines and there are two separate calibration-derivation pipelines: one to generate bias images and another to generate flat-field images. The primary products generated by each pipeline, references to their relevant subsection below, and the frequency at which they are generated are summarized in Table 2.

### 3.1. Data Transfer from Telescope

The ZTF camera generates 16 CCD-based raw data files per exposure of the focal plane (*productID* 1 in Section 4; see also Section 2.1). Prior to packaging by the camera software, the data associated with each exposure (all CCDs) are ∼1.3 GB in size. This includes all exposure, telescope, CCD-image, and overscan metadata. The camera software digitizes its native output into 16-bits per pixel. Furthermore, exposures are separated by typically 40 sec (30 sec integration $+\simeq 8$ sec readout $+\simeq 2$ sec overhead).

The individual CCD-based data files are first transmitted from the P48 to the San Diego Supercomputing Center (SDSC) via the HPWREN[28] microwave network. This link is designed to connect remote areas that are not well served by other technologies. Palomar Observatory is the largest user of

---
[28] http://hpwren.ucsd.edu

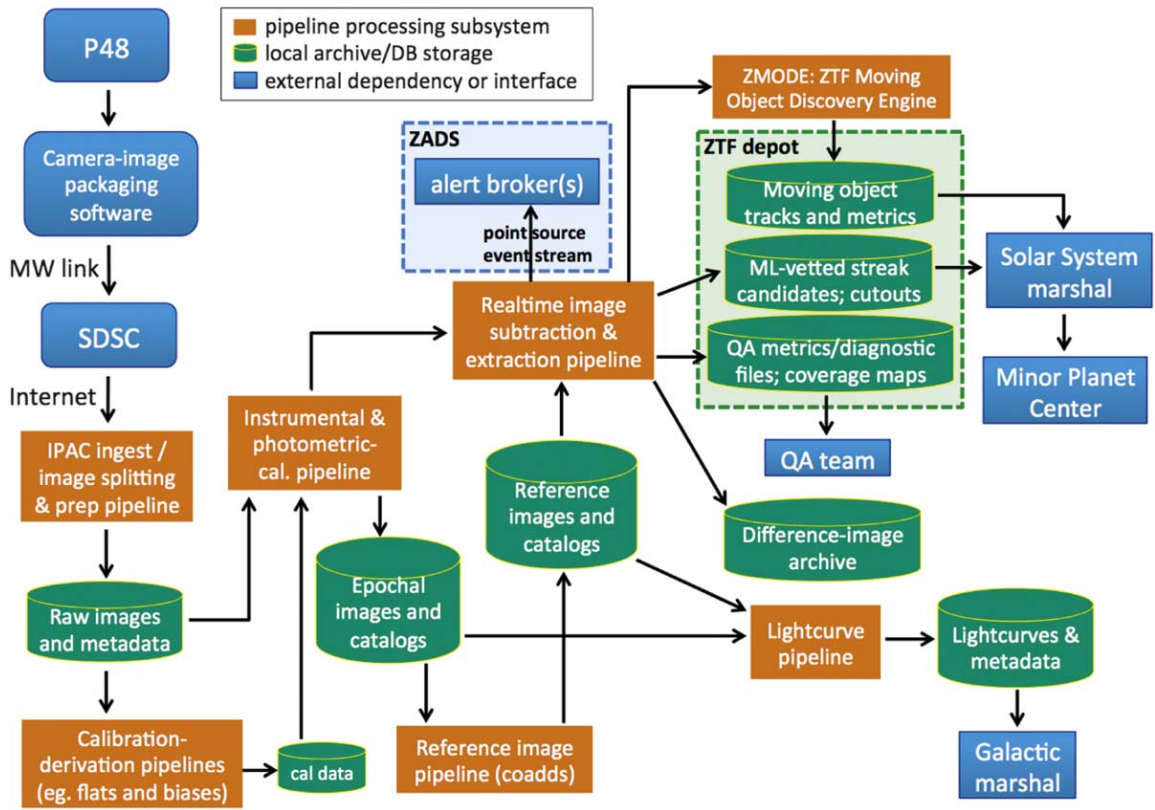

**Figure 2.** Overall data and processing flow in the ZSDS. Green components indicate storage in the form of an archive and/or database residing at IPAC to serve either the public or internal (private) collaboration. Specific products are summarized in Section 4. Vermillion (light brown) components represent the core pipelines, and are described further in Section 3. Blue components indicate external interfaces or dependencies. Acronyms are defined in Appendix A.

Table 2
Data Products

| ProductID | Description | Pipeline | Format[a] | Generation Frequency[b] | Public | Access[c] |
|---|---|---|---|---|---|---|
| 1 | Raw images | Section 3.2 | `fpack`'d MEF | Realtime | Yes | Archive: GUI or API |
| 2 | Epochal science images | Section 3.5 | FITS | Real time | Yes | Archive: GUI or API |
| 3 | Epochal source catalogs | Section 3.5 | FITS binary tables | Real time | Yes | Archive: GUI or API |
| 4 | Point-spread functions | Section 3.5 | FITS, ASCII | Real time | Yes | Archive: GUI or API |
| 5 | Epochal-difference image files | Section 3.6 | `fpack`'d FITS | Real time | Yes | Archive: GUI or API |
| 6 | Reference image (co-add) files | Section 3.7 | FITS | Real time (static) | Yes | Archive: GUI or API |
| 7 | Reference image source catalogs | Section 3.7 | FITS binary tables | real time (static) | Yes | Archive: GUI or API |
| 8 | Calibration image files | Section 3.4 | FITS | Daily | Yes | Archive: GUI or API |
| 9 | Point-source alert packets | Section 3.6 | Avro$^{TM}$ | Real time | Yes | See Section 4.1 |
| 10 | Light curves[d] and metrics | Section 3.8 | Various (selectable) | Updated monthly | yes | Custom GUI and API |
| 11 | Source *matchfiles* | Section 3.8 | Pytable (HDF5) | Updated monthly | No | Datastore for #10 |
| 12 | Streak data (fast-moving SSOs) | Section 3.6 | ASCII, JPEG, ADES | Real time | No | MPC |
| 13 | Moving-object tracks | Section 3.9 | ASCII, JPEG, ADES | Daily | No | MPC |
| 14 | QA metrics and sky-coverage maps | All | ASCII, PNG, MOV, FITS | Real time and daily | Subset | Archive metadata |

**Notes.**
[a] A more detailed description of the file contents, structure, and guidance for use are given in the ZSDS Explanatory Supplement (referenced in Section 1).
[b] Products intended for public release will only be available at public release time (Section 1), and not as soon as they are generated in the ZSDS.
[c] Archive access and tools for analysis and visualization are described in Section 5.
[d] Based on epochal (undifferenced) PSF-fit photometry only (*productID* 3). These are queried from the intermediate *matchfile* products (*productID 11*).

bandwidth on this network. At the start of commissioning, the HPWREN bandwidth was rated at $\lesssim 200$ Mbit/s following overheads and variability (see below). If the raw camera data were immediately transmitted following acquisition (see above), a bandwidth of at least 1.3 GB/40 sec $\simeq 260$ Mbit/s would be required to transfer all data bits in real time. Furthermore, this bandwidth would need to be sustained over the course of a night. Therefore, to accomodate the HPWREN bandwidth with some margin for variability, the raw CCD data files are first compressed by factors of $\approx 2$ to 2.5 at the P48. This compression is slightly lossy and based on the Rice method as implemented in the `fpack`[29] utility. The loss in information due to compression is negligible.[30] Following transmission to the SDSC, the files are pushed to IPAC over the internet.

The performance of the microwave link varies with other usage of the network and weather, particularly if there is an inversion (cloud) layer below the altitude of the P48. Long-term monitoring has shown a median transfer time of a complete exposure of $\simeq 20$ sec with 90th percentile around 24 sec and 99th percentile at 40 sec. Therefore, exposures are typically transferred to IPAC within the observing cadence without any backlog.

### 3.2. Raw Data Ingestion

Upon reception of the raw CCD-based files at IPAC, their filenames and headers are checked for correct formatting and that all the required metadata to facilitate downstream processing are present and within range. The CHECKSUM and DATASUM keywords are also verified in all FITS HDUs. If any of these checks fail, the CCD data file is tagged using a special status flag for future traceability and declared unusable. Such files are not copied to the archive. If a file passes all checks, it is copied to the long-term archive, additional image/exposure identifiers computed, and relevant metadata from the primary header, along with its archived path/filename are stored in the operations database.

At the time of writing, the ingest failure rate is 16 CCD files (or one exposure worth of image data) per night. These are due to timing glitches in the observation start time relative to the "shutter open" time. This is understood and a solution is in progress.

### 3.3. Decompression, Image Splitting, and Preparation

Following ingestion into the archive and operations database, the raw CCD file is decompressed using the `funpack` utility. This creates a multi-extension FITS (MEF) file with pixel values converted from 16-bit integers to 32-bit floating point. The MEF file is then split to extract the four light-sensitive quadrant images together with their headers. The primary HDU containing exposure-level information is prepended to the header of each quadrant image so it can be propagated to all derived image products downstream.

The four corresponding bias-overscan images are also extracted from the MEF file. Each overscan image is used to compute a floating-bias correction using a second-order polynomial fit. This correction is then applied to the respective light-sensitive quadrant image. Statistics are computed on the split quadrant images, their bias-overscans, and polynomial-fit residuals. These metrics are stored in the operations database and the overscan-corrected quadrant images are copied to the pipeline operations file system for later retrieval. All raw CCD files received at IPAC, which includes data intended for calibrations (i.e., biases and flat fields) and engineering-related testing are split into individual quadrant images and overscan-corrected. These preprocessed intermediate quadrant images are fed into downstream pipelines and not copied to the long-term archive.

### 3.4. Calibration Image Generation

There are two separate pipelines that generate calibration image products for use downstream: one to process and stack (zero-exposure) bias images, and another to process and stack exposures of an illuminated "flat screen" to generate relative pixel-to-pixel responsivity maps (high-spatial frequency flat-field images). All processing and stacking occurs at the CCD-quadrant level.

#### 3.4.1. Bias-image Generation

The bias-image generation pipeline takes a list of overscan-corrected zero-exposure bias images corresponding to a specific CCD quadrant. These are acquired prior to each night's observing. A minimum of 20 bias images are required before triggering this pipeline. The bias images are stacked (collapsed) using a trimmed-average per pixel. The trimming is based on $n\sigma$ clipping where $\sigma$ is a robust prior estimate based on percentiles. Five image products are generated: the primary bias-image calibration product; a stacked standard-deviation image $A$ (with pixel outliers removed); a stacked standard-deviation image $B$ (before outlier rejection to support additional masking below); a $1\sigma$ uncertainty image ($=StdDev_{imgA}/\sqrt{N_{trimmed}}$, where $N_{trimmed}$ is the number of samples remaining); and an image of the number of surviving pixels in the stack following outlier rejection. Stacked standard-deviation image $B$ is then thresholded to find and tag "noisy" pixels in a mask image for propagating downstream. The bias-image calibration products are archived and QA metrics are stored in the operations DB.

---
[29] https://heasarc.gsfc.nasa.gov/fitsio/fpack
[30] A consequence of the Rice compression is an increase in the overall pixel noise due to "quantization error." For our compression ratios, the noise increase is $\lesssim 0.2\%$.

### 3.4.2. High-frequency Flat-field Image Generation

A high-frequency flat field is a calibration image whose purpose is to correct for relative pixel-to-pixel responsivity variations, i.e., on spatial scales of a single pixel. The corrections are relative to unity, where unity represents no correction. This image is divided into the science image data by the instrumental calibration pipeline (Section 3.5).

The high-frequency flat generation pipeline takes a list of overscan-corrected images for a specific CCD quadrant and filter corresponding to exposures of an illuminated "flat screen" inside the dome. These are also acquired prior to each night's observing, and are intended to catch temporal variations in the responsivity pattern over the focal plane from night to night. Variations of up to 2.5% (in gradient on CCD-quadrant scales) have been observed across nights. This behavior is not yet fully characterized. Catching and correcting these variations is pivotal because they impact the achieved photometric precision (Section 7.3) as well as gain matching prior to image differencing downstream. A minimum of 20 images is required before triggering this pipeline. Also required is a bad-pixel mask recording the location of prior-masked bad pixels and the bias-calibration image with its accompanying dynamic mask from above (Section 3.4.1). Each input image is preprocessed to subtract the bias image and then pixel values are normalized to a spatially trimmed global mean of one. The trimming of pixel outliers is based on $n\sigma$ clipping. Masked pixels are omitted when computing this mean.

Following preprocessing, the images are stacked (collapsed) using a trimmed-average per pixel. This trimming is also based on $n\sigma$ clipping, where $\sigma$ is a robust prior estimate based on percentiles. The following intermediate image products are generated: a trimmed stack-average; a stacked standard-deviation image $A$ (with pixel outliers removed); a stacked standard-deviation image $B$ (before outlier rejection to support additional masking below); a $1\sigma$ uncertainty image ($=StdDev_{imgA}/\sqrt{N_{trimmed}}$, where $N_{trimmed}$ is the number of samples remaining); and an image of the number of surviving pixels in the stack following outlier rejection. Stacked standard-deviation image $B$ is then used to find and tag any additional "noisy" pixels, i.e., not caught by the bias-image generation pipeline. The pixels in the trimmed stack-averaged image are normalized to a spatially trimmed global mean of one. The resulting image is the final flat-field image (or relative pixel-to-pixel responsivity map) to be used downstream. This image and its ancillary products are archived and QA metrics are stored in the operations DB.

At the time of this writing, an effort is underway to generate and characterize low-spatial frequency responsivity maps by binning the photometric residuals from calibrated source photometry. These maps will be used to validate the high-frequency maps and possibly refine them.

### 3.5. Instrumental Calibration

Instrumental calibration is the first phase of the real-time processing pipeline. Inputs are: a single overscan-corrected CCD-quadrant image in a given filter (Section 3.3); a prior bad-pixel mask; closest-in-time (most recent) calibration images (Section 3.4); lists of prefiltered calibrator sources to support astrometric and photometric calibration covering the CCD-quadrant footprint; and processing parameters. The primary products from this pipeline are *productIDs* 2, 3, 4, 14 in Table 2 (Section 4).

The processing steps are depicted in Figure 3. In summary, this pipeline applies the calibration image corrections (biases and flats; Section 3.4); it corrects for detector non-linearity; it masks aircraft and satellite streaks (using the CreateTrackImage software described in Laher et al. 2014); it performs astrometric calibration (see Section 3.5.1); it masks ghosts from bright sources; it derives PSFs and generates a PSF-fit photometry catalog using a version of DAOPhot (Stetson 1987) optimized for ZTF. This catalog includes the added benefit of de-blending by fitting to multiple detections simultaneously in a two-pass PSF fit. The pipeline then performs photometric calibration (see Section 3.5.2); generates an aperture-based photometry catalog using SExtractor (Bertin & Arnouts 1996); detects and masks cosmic rays; generates an image of pixel uncertainties; computes image QA metrics and statistics on the source catalogs; assigns an overall quality flag with a bit-string recording the details of processing, and stores these in the operations DB; archives the science image product, accompanying bit mask, and source catalogs.

These products can be immediately used as input to the image-differencing (second) phase of the real-time pipeline if other inputs are available (see Section 3.6). Otherwise, the real-time pipeline will terminate and only the science image and catalog products are archived. Having the instrumental calibration steps in their own standalone pipeline also makes reprocessing or testing flexible, i.e., if only the epochal science products are to be replaced.

### 3.5.1. Astrometric Calibration

The astrometric calibration currently uses astrometric reference sources from the *Gaia* DR1 catalog (Gaia Collaboration et al. 2016), with a plan to update to DR2 in the near future. Given that ZTF acquires exposures over a fixed sky grid, partitioned into overlapping fields approximately the same size as the camera field of view (FOV; Bellm et al. 2019), the astrometric reference sources are pre-partitioned into static table files per field and CCD quadrant. A buffer of two arcminutes is used around each predefined survey field/quadrant while creating each table to account for telescope pointing errors and offsets. These static table files are deposited in the parallelized pipeline operations file system (Section 2) to facilitate fast retrieval. During processing (see below), the *Gaia* sources are filtered according to their *G*-filter magnitudes. Only sources satisfying $12 \leqslant G \leqslant 18$ are used for astrometric calibration. Sources that are saturated on the ZTF detectors are excluded.

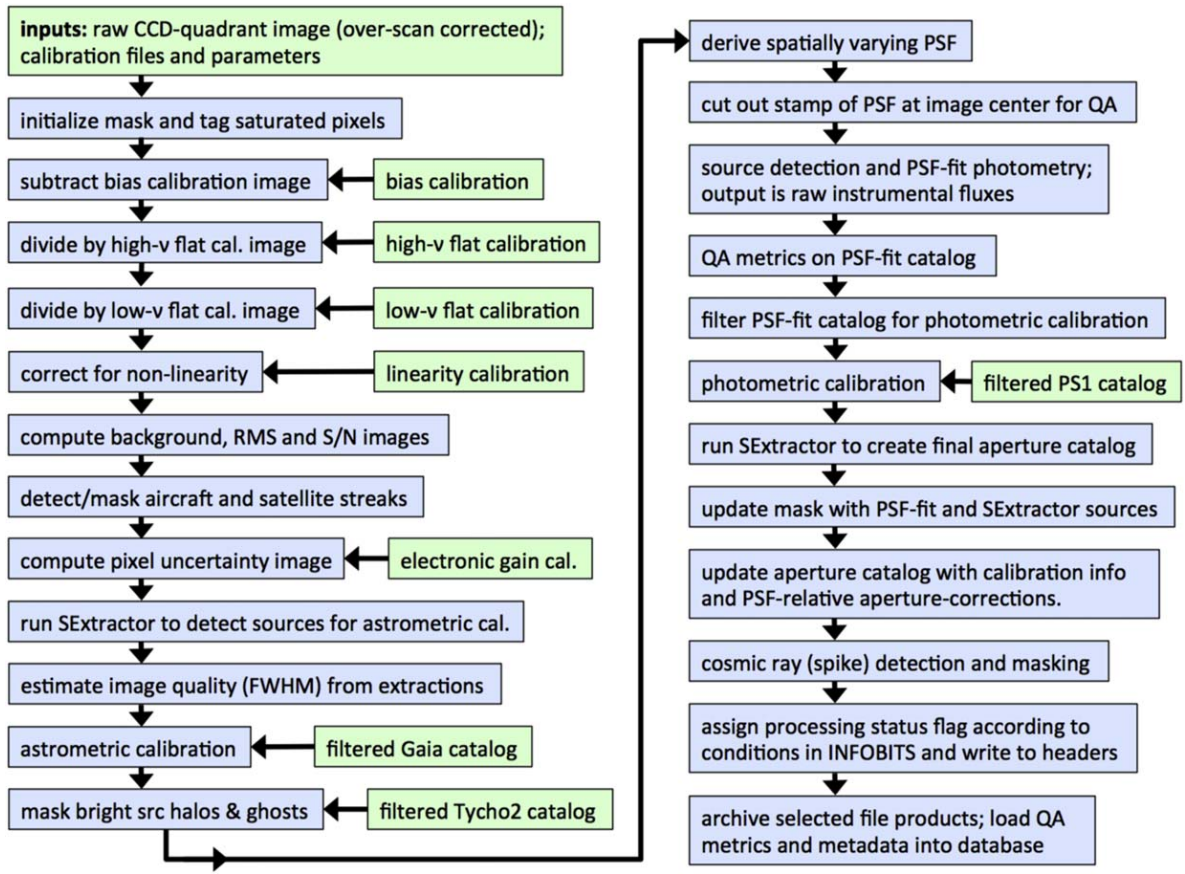

**Figure 3.** Processing flow in the instrumental calibration pipeline. This represents the first phase of the real-time pipeline.

Astrometric solutions are derived using SCAMP (Bertin 2006). SCAMP is executed in two passes. This results in more accurate solutions and makes the process more trackable than solving for all parameters in one run. The first pass uses a distortion prior for the focal plane and initial pointing reported by the telescope control system to transfer the pointing to the center of a quadrant image. This is performed using a pattern match between image extractions (from SExtractor upstream) and astrometric reference sources. This yields refined solutions for the FOV pointing, rotation, and pixel scale at its center. The second pass disables pattern matching and computes a refined fourth-order polynomial distortion solution to replace the prior. The polynomial coefficients are represented using the *TPV* convention.[31] The full astrometric solution (including distortion) is then written to the FITS header of the quadrant image. The input extractions and reference sources are rematched using the final astrometric solution to compute final QA metrics. These metrics are stored in the operations DB. On-sky performance of the astrometric calibration is discussed in Section 7.2.

*3.5.2. Photometric Calibration*

Photometric calibration is performed against a filtered set of calibrator sources selected from the Pan-STARRS1 DR1 (PS1; Chambers et al. 2016) MeanObject database table (Flewelling et al. 2016). The calibrators were selected using criteria to retain primarily photometrically stable stars with reliable and repeatable photometry across multiple epochs of the PS1 survey in each of the *g, r, i, z* filters. Furthermore, to mitigate contamination from source confusion, there are radial-exclusion cuts to maximize the number of isolated stars. As done with the astrometric reference sources (Section 3.5.1), the photometric calibrators are also pre-partitioned into static field/quadrant-based table files in the pipeline operations file system for fast retrieval.

Other inputs to the photometric calibration are sources extracted from the CCD quadrant using PSF-fitting upstream. These are filtered to retain primarily unsaturated stars not contaminated by bad pixels and at some distance from the image edges. The extractions are then positionally matched to the filtered PS1 calibrators. Differences between the instrumental and PS1 magnitudes ($m_{diff} = m_{PS1} - m_{ZTF}$) are filtered for outliers and then used to derive a single calibration zero-point ($ZP_f$) and color term ($c_f$) for the CCD quadrant in ZTF filter $f = g$, $r$, or $i$. The solution is derived from a robust fit of $m_{diff} = ZP_f + c_f\, PS1_{clr}$, where $PS1_{clr} = g_{PS1} - r_{PS1}$, $g_{PS1} - r_{PS1}$, and $r_{PS1} - i_{PS1}$ for $f = g$, $r$, and $i$, respectively. The photometric calibration solution is recorded in the FITS headers of the CCD quadrant and source-extraction catalogs along with uncertainties and quality metrics. Some preliminary results on the performance of the photometric calibration are discussed in Section 7.5.

---
[31] https://fits.gsfc.nasa.gov/registry/tpvwcs/tpv.html

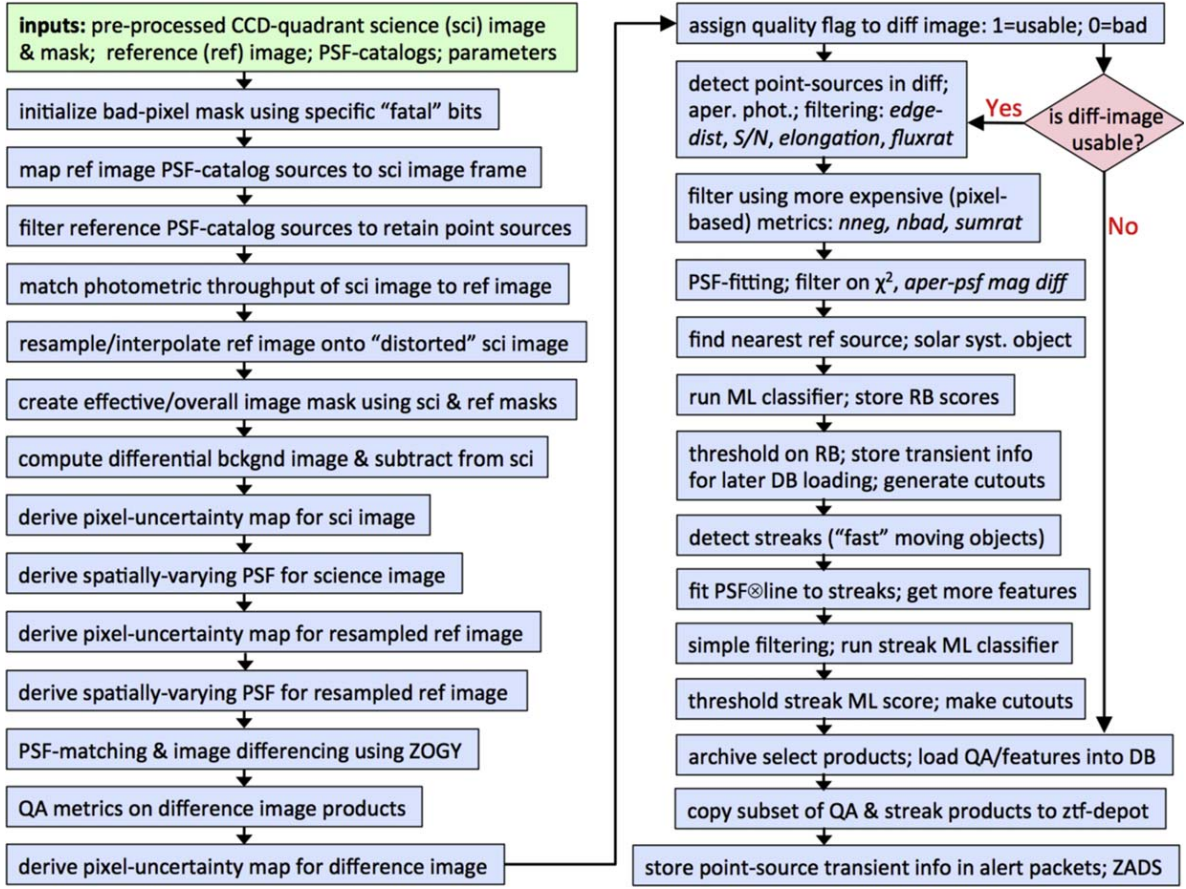

**Figure 4.** Processing flow in the image subtraction/event extraction pipeline, the second phase of the real-time pipeline.

### 3.6. Image Differencing and Event Extraction

Image differencing and event extraction (defined as either candidates for point-source flux transients, SSOs, or streaks from fast(er) moving SSOs) are performed in the second phase of the real-time pipeline, following instrumental calibration (Section 3.5). Image differencing uses as input the epochal CCD-quadrant-based outputs from instrumental calibration, f as well as processing and configuration parameters. It is automatically triggered if a usable (survey-ready) reference image with accompanying PSF-fit photometry catalog (Section 3.7) is available. The primary products from this pipeline are *productIDs* 5, 9, 12, 14 in Table 2 (Section 4). Alert packets (*productID* 9) were described in Section 4.1.

The processing steps are depicted in Figure 4. The first stage of this pipeline consists of preparing the image inputs prior to generating the difference image products. In summary, this step involves using a subset of sources from the input reference and epochal (science) PSF-fit photometry catalogs to match the photometric throughputs of the corresponding images; resamples and interpolates the reference image onto the science image using SWarp (Bertin et al. 2002); masks all bad pixels propagated from the science and reference images; computes a smoothly varying differential background image and subtracts this from the science image. Pixel-uncertainty images and PSFs for the science and reference images are then generated. PSF-matching and image differencing are then performed using the ZOGY algorithm (Zackay et al. 2016). The primary products from this step are the actual difference image; an accompanying match-filtered S/N image optimized for point-source detection; and an effective PSF for the difference image.[32] QA metrics are then derived to assign a usability flag for the difference image. These metrics are thresholded to determine if point-source events should be extracted. That is, the difference image is of sufficient quality as to not strain downstream processing of the event stream due to too many unreliable extractions.

If event extraction is to proceed, the second step of this pipeline detects events from the point-source match-filtered S/N images where detection is performed on both the positive (science minus reference) and negative (reference minus science) images. Note that the negative images are simply $-1\times$ the positive images generated by a single run of the ZOGY software. Events are extracted with both aperture and PSF-fit photometry, and additional source features are computed to support filtering and automated ML-vetting downstream. The events are then lightly filtered to remove obvious false positives, the point-source ML classifier is executed to assign *RealBogus* reliability scores, and image cutouts are generated. Positional association with the nearest known SSO within some specific radius is then performed using the `astcheck`[33] utility and associated metadata for that object is stored. Association with the nearest reference-image-detected source is also performed.

Processing then proceeds with the detection of linear streaks in the positive difference image using the `FindStreaks` software (Waszczak et al. 2017). A simple streak model consisting of a 2D Gaussian representation for the PSF convolved with a line is then fit to each streak (Vereš et al. 2012). This fitting returns estimates for the streak endpoint positions; length; position angle; integrated flux; corresponding uncertainties and covariances; and goodness-of-fit metrics. These metrics are used by the streak ML classifier to assign *RealBogus* reliability scores. These scores are then thresholded to retain likely real candidates from which image cutouts are also generated. The automated ML-vetting steps for both point sources and streaks are described in more detail in Mahabal et al. (2018).

All point-source and streak events, their features, and image QA metrics are stored in the operations DB. The difference image and its PSF are archived and streak metadata with cutouts are pushed to ZTF-Depot for human scanning. A post-processing step gathers all the the point-source event metadata and image cutouts, queries the DB for additional metadata, including cross-matching to the nearest PS1 sources, and then generates an alert packet for each thresholded event (see Section 4.1 for details).

---

[32] In ZOGY's notation (Zackay et al. 2016), these products are respectively $D$, $S_{corr}$, and $P_D$.
[33] https://www.projectpluto.com/astcheck.htm

### 3.7. Reference Image Generation

A reference image provides a static representation of the sky, or more specifically, a historical snapshot as defined by the state of the sky recorded in previous image exposures for a given field, CCD quadrant, and filter. The reference image is a benchmark against which future exposures can be compared (i.e., differenced; Section 3.6) to assist with the discovery of flux transients or moving objects. A catalog of sources extracted therefrom also provides a "seed" for source matching and light-curve generation (see Section 3.8). The primary products from this pipeline are *productIDs* 6, 7, 14 in Table 2 (Section 4).

A reference image is a co-add (a stack-average; see below) of $\geqslant 15$ (minimum) to 40 (maximum) high quality historical CCD-quadrant images. The minimum of 15 was driven by the need to produce as many reference images as possible over the sky early ($\approx$two weeks) into the survey to trigger image differencing and validate the transient-discovery science programs. This minimum was optimal in that it accounted for the availability of "good" quality input images (see below), gave maximal sky coverage, and yielded good quality co-adds usable for image differencing with depths $\simeq 1.5$ mag deeper than the single-exposures. The maximum of 40 was picked to match the image depths from longer camera exposures (up to 300 sec) required by some of the partnership science programs. A large fraction of the reference sky is likely to acquire this maximum depth following bulk regeneration of the reference image library, which has not yet occurred. The maximum of 40 is still somewhat arbitrary and may be higher following a reanalysis of the quality of the current single-exposure image holdings, the quality of the difference-images produced, and transients extracted.

The input images are selected to satisfy a number of quality criteria, primarily images with good astrometric and photometric calibrations, good image quality (point-source FWHM), and robust estimates of pixel noise and background levels falling within specific ranges for each filter. The photometric quality check includes thresholding the calibration zero-point, color term, and limiting magnitude (Section 3.5). Initially, only images satisfying a derived image quality of $2\rlap{.}''0 \leqslant \text{FWHM} \leqslant 3\rlap{.}''5$ are selected. This ensures that most of the inputs are better than Nyquist-sampled (given the $\simeq 1''$ pixel scale). The input images are then ordered according to ascending FWHM and the $N^{th}$ best quality images (where $15 \leqslant N \leqslant 40$; see above) are retained for co-addition. Due to the large number of epochal images from which to draw from, this operation results in most input images spanning a relatively narrow range in FWHM; typically $2\rlap{.}''0 \leqslant \text{FWHM} \lesssim 2\rlap{.}''4$. This restriction in PSF size mitigates inadvertent clipping of pixels containing source signal on and around point-sources in the outlier-trimming step downstream due to the variable PSF across epochs (see below). For more details on the selection criteria for each filter, see the ZSDS

Explanatory Supplement (referenced in Section 1). Other inputs include an accompanying list of bad-pixel masks and processing parameters.

Reference image generation is automated by the pipeline executive (VPO; see Section 2.3) and performed at the end of each observing night. A database query is executed to check which distinct combinations of field, CCD quadrant, and filter which were observed the preceding night, and that do *not already* have an archived reference image, warrant one. That is, do enough good-quality images exist for the given field, CCD quadrant, and filter, accounting for historical images as well? If the minimum number (=15; see above) is satisfied, the reference image pipeline is triggered. The reference image products therefrom are then archived and "locked", i.e., not regenerated unless determined to be of too-low quality for survey use, or until the next round of bulk reference-image regeneration. The input maximum of 40 (see above) pertains to possible future regeneration of reference images by considering all archived historical images.

The first step in the reference-image pipeline consists of regularizing (preconditioning) the input images and creating the co-add image. The science images are first used to determine an optimal co-add footprint geometry and orientation, which defines the final astrometric World Coordinate System (WCS) for the reference image; global median background levels are then subtracted from each respective science image. They are then mapped and resampled onto the co-add footprint using SWarp (Bertin et al. 2002), where distortion is corrected in the process. The science images are then gain-matched because they are very likely to have different photometric throughputs due to variations in atmospheric transparency and the optical system. This is accomplished by rescaling the images using their photometric zeropoints to a fixed target zero-point, one for each filter.[34] The filter-specific target zeropoints become the final zeropoints for the co-add images in each filter, and are stored as MAGZP in their FITS headers.

The resampled and gain-matched images are then combined using outlier-trimmed averaging ($n\sigma$ clipping) on the pixel stacks, with bad pixels (known a priori) removed. The $n\sigma$ clipping is performed on each individual pixel stack where $\sigma$ is a robust estimate of the dispersion in the stack based on percentiles. This process removes the most discrepant outliers, e.g., detector transient artifacts and cosmic rays. Astrophysical transients will also be removed or suppressed, depending on their persistence (longevity) across the input images. Additional pixels in the co-add image are then masked due to possible incomplete (seeing-dependent) saturation masking in the science images upstream. For example, a star with a given magnitude may fluctuate between being saturated at its core when the seeing is good (i.e., most photoelectrons accumulate within a few pixels where the wells have exceeded their capacity), and not saturated when the seeing is bad (photoelectrons are accumulated over a larger region where pixels at/near the source peak have not yet reached full well). QA metrics are then computed on the co-add products and written to their FITS headers. Photometric calibration information is assembled and also written to the headers.

The second step involves constructing source catalogs from the reference image. A spatially varying PSF is first derived for the reference image using a version of DAOPhot (Stetson 1987) optimized[35] for ZTF; sources are detected and PSF-fit photometry is performed (also using DAOPhot) to generate the PSF-fit catalog; aperture and isophotal photometry is then performed using SExtractor (Bertin & Arnouts 1996) and measurements are stored in a separate catalog; aperture (curve-of-growth) corrections relative to the PSF-fit photometry are then computed for the fixed aperture measurements using a subset of filtered sources; these corrections are stored in the header of the aperture catalog table; QA metrics are computed from the catalogs and photometric calibration information is written to their headers.

An overall quality status flag is assigned to the reference-image (and associated products) by thresholding on specific QA values and stored in the operations DB for bookkeeping. The primary reference image, depth-of-coverage map, pixel-uncertainty image, and catalog table files are copied to the archive with paths/filenames registered in the operations DB for later retrieval by the image differencing (Section 3.6) and source-matching (Section 3.8) pipelines.

It is important to note that reference images are not recalibrated against PS1 using its PSF-fit catalog to assign new zeropoints and color terms, i.e., as done for epochal images in the instrumental calibration pipeline (Section 3.5). This is because we rely on the accuracy of the photometric calibration solutions from the input images, which were nonetheless preselected to be of high quality. Throughput matching of the input images to a common fixed zero-point handles this implicitly. Furthermore, the final reference-image WCS (defining the co-add footprint; see above) is not astrometrically recalibrated or refined. The input images were also preselected to have good astrometric quality. No relative astrometric refinements are needed prior to co-addition.

### 3.8. Source Matching and Photometric Refinement

The purpose of the source-matching pipeline is to construct light curves from the multi-epoch photometry catalogs generated by the instrumental calibration pipeline (Section 3.5). Only the PSF-fit photometry catalogs corresponding to a single field, CCD quadrant, and filter are positionally cross-matched across epochs, going back to the start of the science survey. The starting epoch is configurable. The primary product from this pipeline is *productID* 11 in Table 2 (Section 4), a product referred to as a *matchfile*. This is a file in HDF5 format[36] containing all the light curves and their statistics for all matched sources in the footprint defined by a specific field, CCD quadrant, and filter. The contents of these matchfiles are extracted by the light-curve query service (Section 5.3 or *productID* 10 in Table 2).

---

[34] The choice is arbitrary. To avoid large or small floating point values in the pixel data following rescaling, the target zeropoints were derived to be ≈close to the input image calibration zeropoints by medianing the zeropoints of a few hundred thousand images. The derived zeropoints are fixed for the entire survey.

[35] The major improvement is reducing memory use and overall runtime.

*Matchfile* creation and updating of existing versions as data are accumulated is performed on timescales of one month or longer because it is contingent on having enough (new) epochal data in a specific field, CCD quadrant, and filter. A good-quality (survey-ready) reference image source catalog (Section 3.7) for this footprint is also required. This is used to seed the source-matching process (see below). The source-matching pipeline is scheduled at times when no other pipelines are running; typically during daylight hours.

In summary, the steps used to create a *matchfile* are as follows. The HDF5-structured tables and columns referenced below are defined in the ZSDS Explanatory Supplement (referenced in Section 1). If a matchfile is to be created for the first time, a PSF-fit source catalog corresponding to the reference image is stored. PSF-fit source catalogs corresponding to the epochal images are also stored. The reference image catalog is first used to seed the creation of a *sources* table in the matchfile. The following empty HDF5 tables are then created in the matchfile: *exposures*, *sourcedata*, *transients*, and *transientdata*. A positional match to the coordinates in the *sources* table is performed and for each epochal catalog, a single entry is made in the *exposures* table. Each source that matches a reference source is assigned the *matchid* from the *sources* table. Photometry for that source is then entered in the *sourcedata* table. Any input detection that is not matched to a reference source is considered a transient. A position match is made against coordinates in the *transients* table. Transients receive their own *matchid* and their photometry is recorded in the *transientdata* table. If alternatively a pre-existing matchfile is to be updated with new data, its path/filename is required on input. In this case, the new epochal catalogs are positionally matched to the existing sources and their measurements appended to the matchfile.

Following the source-matching, the epochal photometry can be optionally refined in a relative sense. The default is to perform relative photometry prior to delivery of a matchfile to the archive to support light-curve queries (Section 5.3). Relative corrections to the input epochal photometric zero-points (Section 3.5) are computed based on a maximum of 5000 stable (least-variable) sources in the given CCD-quadrant footprint. These corrections are computed using the algorithm described in Ofek et al. (2011). See also Levitan (2013). A *delta-zero-point* correction is computed for each single-epoch exposure and stored in the exposures table. Both corrected and uncorrected magnitudes are stored for each source in the matchfile product. This relative photometric refinement uses *all* the epochal measurements in the matchfile, going back to the earliest epoch. Therefore, as new data are matched and appended, subsequent reruns of relative photometry will cause all delta-zero-point corrections to slightly shift. Lightcurve measurements will therefore also shift following application of the new corrections. The performance of relative photometric refinement is presented in Section 7.3.

Following source-matching and (optionally) relative photometric refinement, collapsed-light-curve statistics for sources and transients are computed, and stored in their respective HDF5 tables in the matchfile. These statistics include median and mean magnitudes, dispersion measures (RMS and percentile-based), skewness, kurtosis, chi-square, minimum and maximum, mean pairwise slope, number of measurements below/above various RMS cuts, Stetson indices, and more. These statistics enable users to query light curves exhibiting specific properties or shapes.

The matchfiles are copied from the pipeline operations file system to the archive. The archive ingestion process also reads the *matchids*, associated time-collapsed statistics, metadata, and reference-position information for all light curves and stores these in an "objects" database. Actual light curves (the single-epoch photometry points) are not stored in this database. The objects database interfaces with a GUI to enable spatial searches and range queries on the pre-computed statistics (Section 5.3). The returned object IDs are used to retrieve the light curves from the associated matchfile(s). The light curves are then packaged and delivered to the user.

### 3.9. Moving Object Track Generation

A moving-object track is a collection of linked *point-source* events extracted by the image-differencing pipeline (Section 3.6) corresponding potentially to a SSO. This includes NEAs. Only difference-image point-source events detected across multiple exposures are linked, not streaks produced by faster moving objects. The primary software used to construct candidate moving-object tracks is called ZMODE: ZTF's Moving Object Discovery Engine. The products from ZMODE were summarized under *productID* 13 in Table 2 (Section 4). A detailed listing of the specific file products is given in the ZSDS Explanatory Supplement (referenced in Section 1).

Below are the high-level steps carried out by the ZMODE pipeline executive script (managed by the VPO; see Section 2.3). The actual track-generation process is described in Appendix B. The ZMODE executive is triggered at the end of each observing night using point-source extractions spanning at most the last four continuous observing nights if available. If four continuous observing nights are not available, detections from either the most recent three or two nights spanning a four day interval are used as input. The primary inputs are a list of the night dates from which the difference-image-based point-source extractions should be queried, thresholds for the input detections (e.g., ML-*RealBogus* score, S/N, and magnitude limit), velocity-matching thresholds, and thresholds for filtering the orbit-fit quality metrics. A database query is executed to retrieve the point sources satisfying the input criteria. The `ZMODE` software is then executed on this list of sources to construct candidate moving-object tracks (see Appendix B). Orbit fitting is then performed on each candidate track using the `find_orb`[37] software. The output metrics from `find_orb` are thresholded to construct a list of likely real moving-objects, i.e., worthy for downstream human vetting prior to submission to the Minor Planet Center (MPC). An orbital elements file is also generated.

---
[36] http://portal.hdfgroup.org/display/HDF5

Following orbit fitting and filtering, the following products are generated: a provisional MPC report with object ephemerides in the traditional 80-character format; a report following the new Astrometry Data Exchange Standard (ADES[38]) format; a report containing scores for the probable orbit classes for each object generated by the `digest2`[39] software; science and reference image cutouts from the images used in image differencing, centered on every detection used to construct each *new* track (i.e., not associated with a known SSO). The MPC report files include matches to both previously known objects and potentially new objects not recovered from cross-matching to the MPC archive. All products are copied to ZTF-Depot for human vetting.

## 4. Data Products

Table 2 summarizes the primary data products, formats, generation frequency in the ZSDS, whether the product is destined for future public release, and access portal(s). The *Pipeline* column refers to the section that described the relevant pipeline that generated the product. Acronyms are defined in Appendix A. As mentioned earlier (Section 2.1), all products (except *productIDs* 1, 10, and 13) are generated per filter and CCD quadrant. Below we describe each product according to its *productID* in Table 2. More detailed descriptions and usage instructions can be found in the ZSDS Explanatory Supplement (referenced in Section 1).

1. *Raw image files.* These are `fpack`-compressed Multi-Extension FITS[40]-formatted files containing all the quadrant images and their bias (overscan) strips for a single CCD from an exposure. Each quadrant image resides in a separate extension with a header. There are a total of nine extensions with the first containing generic header metadata for the exposure. There are 16 separate CCD data files per camera exposure.

2. *Epochal science image files.* These are instrumentally calibrated CCD-quadrant based images, where each pertains to a single camera exposure. Their headers contain astrometric and photometric solutions, quality information, and other metadata. Accompanying these are 16-bit mask images reporting the locations of bad pixels with specific bits assigned to different conditions. These conditions are defined in their headers. Processing log files are also provided.

3. *Epochal source catalogs.* Accompanying each epochal science image are two source catalogs: one containing PSF-fit photometry only, and another with measurements using exclusively fixed concentric circular apertures and adaptively determined isophotal photometry. Source-shape metrics (e.g., ellipticities and second-order moments) are also included. Each catalog is treated independently and seeded by different source detection lists. It is important to note that the PSF-fit catalogs are optimal for point-source photometry only while the aperture catalogs are more generic and contain measurements that are useful for extended sources. The metrics in the aperture-based catalogs are based on `SExtractor` (Bertin & Arnouts 1996) and were inherited from the PTF survey products. As described in Section 3.5.2, photometric calibration is exclusively derived using a filtered subset of point sources from the PSF-fit catalog cross-matched to PS1. The zero-point and color-term solutions are therefore optimized for photometric measurements of *point sources* in both the PSF-fit and aperture catalogs. The calibration solution is written to the FITS headers of both catalogs. Furthermore, aperture (curve-of-growth) corrections (relative to the accompanying PSF-catalog measurements) are computed for each fixed circular aperture and written to the FITS header of the aperture catalog.

4. *Point-spread functions.* PSF template used to generate the PSF-fit photometry catalog (*productID* 3). This comprises of a FITS-formatted image of the PSF represented at the center of the epochal image, and an ASCII file (in `DAOPhot`'s LUT format; Stetson 1987) containing estimates from a linearly spatially varying model fit.

5. *Epochal-difference image files.* Difference image constructed from preregistered, gain-matched, and PSF-matched science and reference images (inputs from *productIDs* 2 and 6) in the sense: science minus reference. This is in `fpack`-compressed[38] FITS format. An estimate of the PSF for this difference image in FITS format is also provided. This represents an image of the effective PSF for the entire (CCD-quadrant-based) difference image, as the estimation process

---

[37] https://www.projectpluto.com/find_orb.htmf
[38] https://minorplanetcenter.net/iau/info/ADES.html
[39] https://godoc.org/github.com/soniakeys/digest2
[40] https://fits.gsfc.nasa.gov

6. *Reference image (co-add) files*. These co-adds are files of single-epoch science images from two satisfying a number of quality criteria (primarily with "good" astrometric and photometric calibration and image quality). Accompanying the primary reference image is a pixel depth-of-coverage image representing effectively the number of input pixels used (accounting for losses due to bad pixels, saturation, and outliers), and an uncertainty image storing the $1\sigma$ uncertainty in pixel flux. The headers contain astrometric and photometric calibration solutions, and QA metadata. Processing log files are also provided.

7. *Reference image source catalogs*. Accompanying each reference image product are two source catalogs: one containing PSF-fit photometry only, and another with measurements using exclusively fixed concentric circular apertures and adaptively determined isophotal photometry. Source-shape metrics (e.g., ellipticities and second-order moments) are also included. The contents and formats of both the PSF and aperture-based catalogs are similar to those in *productID* 3, and were generated using the same software (Section 3.7).

8. *Calibration image files*. These comprise bias-image calibration maps, relative pixel-to-pixel responsivity maps (or high-frequency flats), accompanying $1\sigma$ pixel-uncertainty maps, bad-pixel masks that tag outliers identified from image stacking, and processing logs.

9. *Point-source alert packets*. See Section 4.1.

10. *Light curves and metrics*. These are light curves generated from positionally matching sources detected across epochs from the PSF-fit photometry catalogs (*productID* 3). The reference image PSF-fit photometry catalog (*productID* 7) is used to seed the position matching. The source-matched photometry is further refined in a relative sense across epochs per CCD quadrant. Metrics for each light curve (global statistics from collapsing measurements across all epochs) are also available. A custom GUI and API (Section 5) allows one to retrieve light curves using cone-searches and constraints on any of the metrics.

11. *Source matchfiles per-image*. These are internally generated intermediate files in *Pytable* (HDF5) format where each contains all the light curves from positionally matching sources across epochs from the PSF-fit photometry catalogs (*productID* 3) falling in a single field, CCD quadrant, and filter. These products feed the light-curve query service (*productID* 10) and are not available as a standalone product from the archive.

12. *Streak data (fast-moving SSOs)*. These are products to support discovery and characterization of SSOs (NEAs specifically) that are moving fast enough to streak in the individual CCD-quadrant exposure images. These are detected in the *science minus reference* difference images. Products consist of QA metadata, streak diagnostics (integrated flux, width, length, end-point positions, ML-vetted scores), and image cutouts on the science and reference images. Products are copied to ZTF-Depot for human vetting. Likely real candidates are delivered to the MPC in the ADES (XML-based) format. Once delivered and ingested, they can be accessed from the MPC[41] for follow-up by the community.

13. *Moving object tracks*. These consist of linked point-source events detected from the *science minus reference* difference images spanning a number of consecutive nights that correspond (potentially) to the same moving SSOs. Each track consists of ephemeral astrometry and photometry for each detection, and is accompanied by QA metadata from orbit fitting, associations with known objects, and image cutouts on each detection. These products are copied to ZTF-Depot and are only internally accessible to specific project members for vetting. Likely real candidates are delivered to the MPC in the ADES (XML-based) format. Once delivered and ingested, they can be accessed from the MPC.[41]

14. *QA metrics and sky-coverage maps*. These are image quality and calibration performance metrics, and statistics on the number of extracted sources are generated by all pipelines. This includes quality metrics on the raw image data (*productID* 1) as well as telemetry from the observing system and telescope extracted from the raw data files. Furthermore, *Aitoff*-projected survey depth-of-coverage maps are generated at the end of each observing night per filter. Both cumulative and nightly (incremental) coverage maps are produced in a number of coordinate systems. Movies of the cumulative maps are also available. All QA metrics and coverage maps are copied to ZTF-Depot to support internal analyses. A subset of the QA metrics accompany the archived products. Survey depth-of-coverage maps will be made available to the community at each scheduled public data release.

### 4.1. Alert Packets

Alert packets are binary files containing metadata and contextual information for a single "event" extracted from either a science minus reference difference image or its negative, a reference minus science difference image (see Section 3.6). Such an event may have been triggered from a flux-transient, a reoccurring flux-variable, or moving object. These raw events are lightly filtered to remove obvious false positives, and only those remaining are packetized for distribution. This filtering is discussed in the ZSDS Explanatory Supplement (referenced in Section 1) and in Mahabal et al. (2018).

---

[41] https://www.minorplanetcenter.net/iau/mpc.html or specifically the NEOCP: https://www.minorplanetcenter.net/iau/NEO/toconfirm_tabular.html.

Alert packet files are in the Apache Avro[TM] format[42], a binary serialization format for efficient distribution. Alert packets are the basic atomic units of the ZTF Alert Distribution System (ZADS; Patterson et al. 2019). This system comprises staging of alert packets in the ZSDS Kafka (see footnote 6) cluster at IPAC (the producer), mirrors with Kafka clusters located elsewhere, in this case UW, interfaces with other consumers in the consortium, and interfaces with the "alert brokers" (e.g., Narayan et al. 2018).

The information in an alert packet is structured using JSON-based schemas. A description of all the metadata can be found in the ZSDS Explanatory Supplement (referenced in Section 1). Here is a summary of the content:

1. An alert name (or *objectId*). The same *objectId* can correspond to multiple alert packets, for instance, that are triggered on the same astrophysical object detected across distinct observational epochs. That is, a previously assigned *objectId* (going back to the beginning of the survey) is reused on any new alert falling within 1.5 arcsec of that same sky position.
2. Source-specific features (metrics) for the difference-image-detected event that triggered the alert. These features facilitate additional ML-vetting downstream.
3. Image-specific metadata (also features) for the corresponding science, reference, and difference images. These comprise quality metrics from image differencing, observational span of the inputs used to construct the reference image, limiting magnitudes, and image identifiers to facilitate retrieval of ancillary products from the archive. The metrics also include a count on the number of previous candidate events detected near/on the alert position and the number of epochal images covering (touching) that position. This metadata can also be used to guide additional ML-vetting.
4. The closest known solar-system object falling within a specific radius. If found, its angular separation, name, and *V*-band magnitude (if available) are reported.
5. The closest, second closest, and third closest PS1 source falling within a specific radius. If found, angular separations, magnitudes, PS1 catalog IDs, and ML-predicted star-galaxy classification scores for all PS1 matches are reported. The PS1 star-galaxy classification process is described in Tachibana & Miller (2018).
6. The closest source from the *Gaia* DR1 catalog, including the brightest source below some magnitude limit within a specific radius. If found, angular separations and *Gaia* *G*-band magnitudes are reported.
7. Previous (historical) events falling within 1.5 arcsec of the alert position (if any) with source features and image-based metrics for each event. The search span for possible associated historical events is currently 30 days. These historical events have more relaxed extraction criteria than those used to trigger the initial alert. These include detections from all filters and from either positive (science minus reference) or negative (reference minus science) difference images. If the alert position touched a previous image but was not associated with a detected event (from either positive or negative difference), a flux upper-limit and "null" entries for the other metadata are reported instead. The flux upper-limit is based on a global difference image estimate.
8. Three 63 × 63 native pixel image cutouts centered on the alert position from the science, reference, and difference images. These cutouts are in lossless *.fits.gz* format with pixel values retaining the full 32-bit floating point precision inherited from the full input images.

Details on how to access the alert packets in near-real-time and supporting documentation are available on the ZTF public website.[18] Furthermore, all alert packets corresponding to a CCD-quadrant are packaged into a tar-gzipped file. These files are then copied to the long-term archive at IRSA along with other quadrant-based science products. See Section 5 for access details.

## 5. Archive Access, Services, and Tools

The ZTF archive, associated services, and supporting documentation can be accessed from the URL referenced in Section 1. Epochal-based science products are copied to the archive in near-real-time (i.e., immediately following generation). Reference images (co-adds) made therefrom are archived soon after they are created. The archived file products are summarized in Table 2 and described in Section 4.

### 5.1. Generic File-based Products

There are two ways to access the image and catalog file products from the ZTF archive.

The first and most common access method is via the GUI.[43] This allows a user to supply a sky position or a list of positions in any coordinate system. Alternatively, one can specify the names of astrophysical objects resolved by either NED[44] or SIMBAD.[45] Searches can also be performed by survey field ID and optionally CCD ID therein. The query returns a list of the CCD-quadrants touching those positions with associated observational metadata. This metadata (including observation timestamps) can be filtered to retain a subset of products for later packaging and downloading. All or a subset of the ancillary products per CCD-quadrant can be downloaded, including pre-packaged alert packets (Section 4.1). The results page also displays image previews for selected CCD-quadrants. These can be explored and manipulated interactively in the browser. This includes the ability to compute image statistics within sub-regions, perform cutouts and save them, or overlay sources from catalogs available in IRSA, from a VO[46] registry, or from a custom upload.

---

[42] https://avro.apache.org
[43] https://irsa.ipac.caltech.edu/applications/ztf/
[44] https://ned.ipac.caltech.edu
[45] http://simbad.u-strasbg.fr/simbad/

The second access method is via an API.[47] This method is non-interactive and useful for executing queries and downloads from within user-customized scripts using either `wget` or `curl` commands. The process involves first issuing a query using constraints on the available searchable quadrant-image metadata (seeded by sky position). This returns a list of the CCD-quadrants and their metadata in a table. The user can then construct the archive product URL-paths and filenames using the contents of this table for subsequent downloading using scripted `wget` or `curl` calls. Cutouts can also be performed on each image instead.

### 5.2. Solar-system Object Searches and Precovery

The GUI[43] described in Section 5.1 also allows one to specify the name of a SSO, either its official numbered designation, its NAIF ID[48], or its MPC-formatted ephemeris. Alternatively, orbital parameters with a search timespan can be supplied. This returns a list of the CCD-quadrants containing apparitions of the moving object, their metadata, and image previews. The same exploratory services discussed above can then be used.

A more customized tool for exploring and characterizing known SSOs for ZTF is the Moving Object Search Tool[49] (`MOST`). This tool returns more metadata for the object, diagnostic orbit plots, orbital elements, ephemeris information, and links to scripts for downloading archived images containing the object. An accompanying API[50] to perform `MOST` queries using either `wget` or `curl` is also available. This API returns a table of metadata from which the URL-paths and filenames of archive products can be constructed for subsequent downloading. Image cutouts can also be performed on the positions of interest in each epochal image.

### 5.3. Lightcurve Retrieval and Analysis

A separate custom GUI is available to query photometrically refined light curves from pre-positionally matched sources. This is *productID* 10 in Table 2. In essence, this service will allow cone-searches and filtering on any of the available pre-computed light-curve-collapsed metrics. A database of the (reference-image extracted) objects used to seed each lightcurve during source matching (Section 3.8) along with their metrics can be queried standalone. This service will enable one to get an estimate of the number of retrievable light curves over regions of sky that satisfy any filters before querying and downloading them. APIs to perform these tasks are also available.

Connected to the ZTF lightcurve GUI is a separate generic time-series viewer and analysis tool.[51] If the light curves were saved in table files, e.g., from an API call, these can be uploaded to the time-series tool. This tool will ingest the light curves and allow one to display zoomed-in images at each timepoint, interactively manipulate them, compute periodograms, periods, other statistics, edit/remove photometry points, and more.

## 6. Data Volumes, Rates, and Source Statistics

Guided by the accumulated survey data, below are the typical data volumes and statistics produced *per night*. These estimates are based on an average uninterrupted observing night spanning ∼8 hr 40 min.

1. Number of camera (on-sky science) exposures: ∼700
2. Number of calibration exposures: 84
3. Raw incoming data rate: ≃260 Mbit/s (uncompressed)
4. Volume of raw image data: ∼1 TB (uncompressed)
5. Volume of real-time data products: ∼4 TB
6. Number of CCD quadrant image product files (science, difference, mask images, and catalog tables): $\sim 2.3 \times 10^5$
7. Number of unfiltered (unvetted) $5\sigma$ point-source alerts[52] extracted from difference images; includes events from flux transients, variables, and SSOs: $\sim 10^5$ to $2 \times 10^6$ (see also Section 7.6).
8. Number of likely *real* (ML-vetted) $5\sigma$ point-source alerts[52] extracted from difference images; includes events from flux transients, variables, and SSOs:[53] $\sim 10^3$ to $10^5$
9. Number of ML-vetted (but not human-vetted) streaks[52] extracted from difference images as candidates[54] for fast-moving SSOs: $\lesssim 1.5 \times 10^3$
10. Expected number of real (human-vetted) streaks extracted from difference images caused by fast-moving SSOs: $\lesssim 3$
11. Number of single-epoch image PSF-fit source measurements: ∼0.5 to 1 billion (sky location dependent)
12. Number of single-epoch image aperture source measurements: ∼300 to 500 million (sky location dependent)

---

[46] http://www.ivoa.net
[47] https://irsa.ipac.caltech.edu/docs/program_interface/ztf_api.html
[48] https://naif.jpl.nasa.gov/pub/naif/toolkit_docs/C/req/naif_ids.html
[49] http://irsa.ipac.caltech.edu/applications/MOST/
[50] http://irsa.ipac.caltech.edu/applications/MOST/MOSTProgramInterface.html
[51] https://irsa.ipac.caltech.edu/irsaviewer/timeseries
[52] Depends on sky location and available reference image coverage in all filters.
[53] Number can be dominated by asteroids and repeated apparitions thereof within the same night.
[54] Most of these are artificial satellites, unmasked segments of aircraft trails, and unmasked CCD-bleed artifacts from saturated sources.

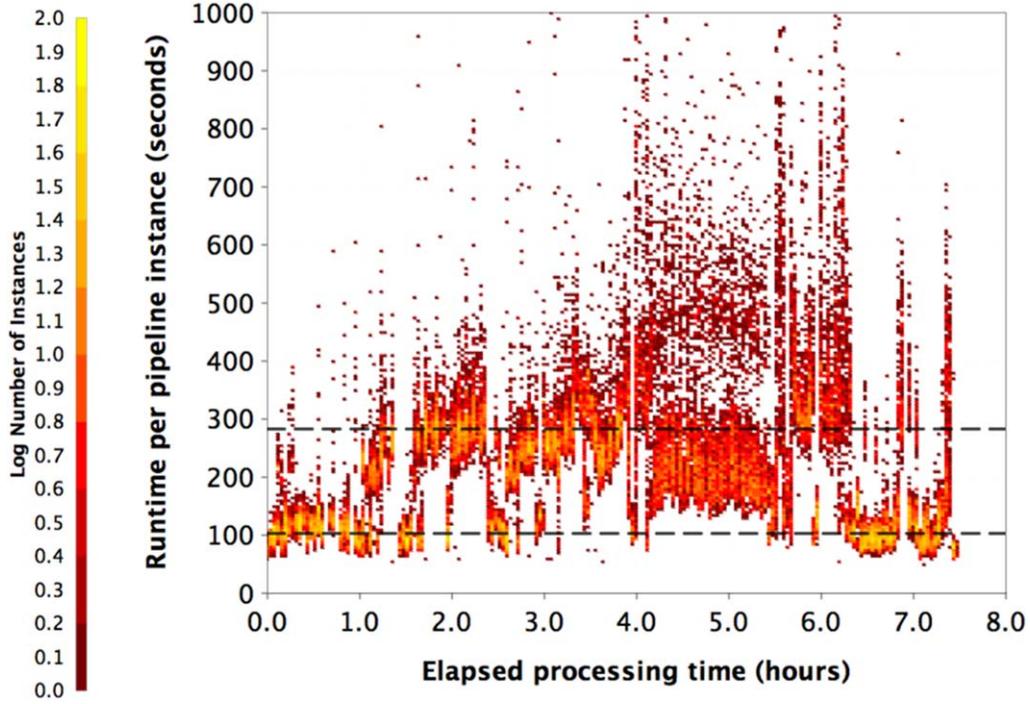

**Figure 5.** Real-time pipeline runtime per "pipeline instance" where an instance corresponds to the processing of a single CCD-quadrant image, end-to-end. Horizontal dashed lines are the median runtimes with image differencing and alert generation included (~290 sec) and without, i.e., only instrumental calibration (~100 sec). See Section 7.1 for details.

To convert the above source statistics to average areal densities in units of $\#/\deg^2$, multiply by $\simeq 3 \times 10^{-5} \deg^{-2}$.

At the end of the nominal *three-year survey* and assuming 260 good-weather observing nights per year, we expect:

1. Volume of all data products: ~3.2 PB (products are summarized in Table 2)
2. Volume of reference image (co-add) products: ~60 TB
3. Number of CCD-quadrant-based reference images in *g, r, i*: ~$2.8 \times 10^5$
4. Volume of *matchfile* products (light-curve files; Section 4): ~50 TB
5. Number of CCD-quadrant based *matchfile*s in *g, r, i*: ~$2.8 \times 10^5$
6. Number of PSF-fit extracted reference image sources used to seed *matchfile* (light curve) creation: ~1.8 billion
7. Number of single-epoch image PSF-fit source measurements: ~800 billion
8. Number of single-epoch image aperture source measurements: ~230 billion

## 7. Pipeline and On-Sky Performance

### 7.1. Realtime Pipeline Runtime

Figure 5 shows a runtime versus elapsed-time density plot for processing a full night's worth of data (specifically night 2018-06-17UT). This plot includes 38,904 independent pipeline instances processed using the 66-node cluster with eight instances per node (see Section 2.1). An instance corresponds to the processing of one CCD-quadrant image through the real-time pipeline, which consists of two sub-pipelines: the instrumental calibration pipeline (phase 1; Section 3.5) and the image-differencing/event-extraction pipeline (phase 2; Section 3.5).

As seen in Figure 5, approximately 60% of the images from this night went through the image-differencing pipeline. This is because not all fields and CCD-quadrants observed had pre-existing reference images in the archive to trigger image differencing. Therefore, this plot compares timing results for phase 1 of the real-time pipeline, which is ~1.5 to 2 minutes per CCD-quadrant, and phases 1 + 2 (full pipeline, including alert packet generation and streak detection), which amounts to ≲5.8 minutes per quadrant for most images observed out of the Galactic plane. The longer runtimes (>360 seconds or >6 minutes) between the elapsed hours of ~4 and 6.3 are due to processing of images in the Galactic plane (at $|b| \lesssim 3.5°$, $3.4° \lesssim l \lesssim 80°$) where the source density is substantially higher (see below). The runtime here can exceed 12 minutes per quadrant (95th percentile). This night spanned ~7 hr 30 min and was processed in approximately the same time, i.e., with no accumulated backlog.

The primary factor that affects pipeline runtime is source density, e.g., in the Galactic plane. This puts a strain on source extraction, in particular PSF-fitting and de-blending, and also on event extraction and alert packet generation downstream. The latter is also taxing in the ecliptic plane due to the large number of SSOs that are repeatedly detected.

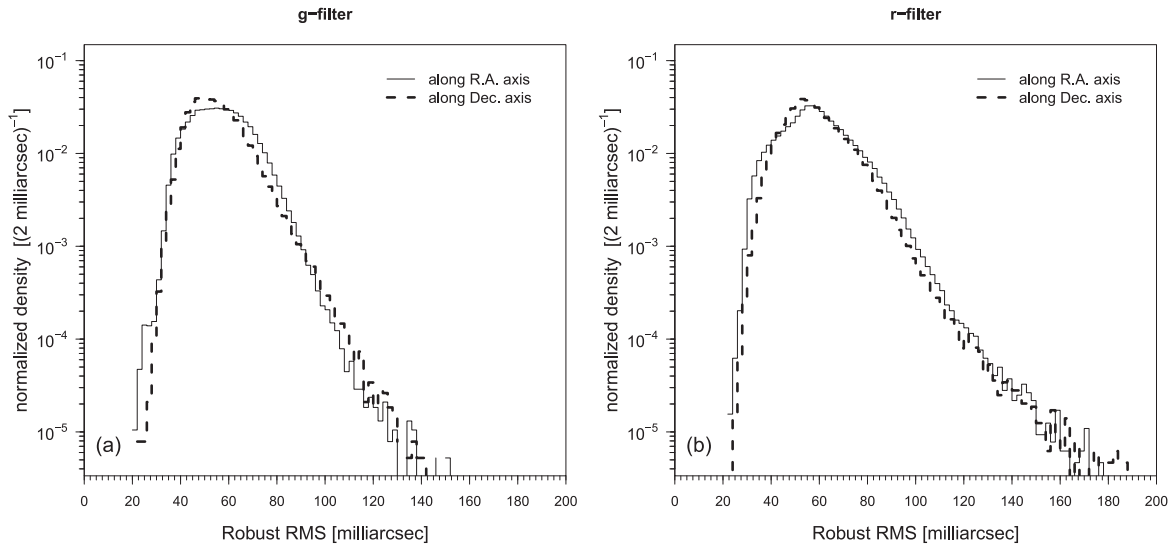

**Figure 6.** Distributions of robust astrometric RMS per processed CCD quadrant along each axis with respect to *Gaia* DR1 for (a) *g*-filter and (b) *r*-filter. The robust RMS is from the percentile difference: (84%–16%)/2. These estimates pertain to ZTF extractions with photometric $S/N \geqslant 10$. See Section 7.2 for details.

### 7.2. Astrometric Accuracy

The astrometric calibration process was summarized in Section 3.5. The accuracy in pipeline-reconstructed astrometry in the *g* and *r* filters at airmass $\leqslant 1.1$ is typically 45 to 65 milliarcsec, measured as the mode in the distribution of RMS errors per axis. Figure 6 shows distributions of the RMS per filter for 190,641 *g*-filter and 321,420 *r*-filter quadrant images acquired from 2018-02-05UT to 2018-04-01UT, and observed only at airmass $\leqslant 1.1$. The astrometric accuracy is slightly better in the *g*-filter due to its better image quality in general (effective point-source FWHM); i.e., the accuracy of source-centroids depends on the overall image quality. For a 2D Gaussian distribution, the RMS in Figure 6 is equivalent to the median radial separation.

It is important to note that the RMS values in Figure 6 are based on the astrometric residuals of extracted source positions with respect to *Gaia* DR1 using ZTF sources with photometric $S/N \geqslant 10$. At this level, errors in background estimation and Poisson noise will dominate the positional uncertainties. Hence, the RMS values reported above overestimate the true achievable astrometric *precision*. Figure 7 shows the RMS as a function of *g*-magnitude for a single CCD-quadrant. From photometric repeatability estimates (Section 7.3), $S/N \geqslant 10$ and $S/N \geqslant 40$ for example correspond to magnitudes of $\lesssim 20$ and $\lesssim 18$ respectively. For magnitudes $\lesssim 18$, the RMS is $\lesssim 30$ milliarcsec per axis at low airmass ($\leqslant 1.1$) and more suitably represents the limiting astrometric precision.

It is also worth noting that the ($S/N \geqslant 10$) modal RMS values reported above are two to four times larger than the positional accuracy of sources in the *Gaia* DR1 catalog (Gaia Collaboration et al. 2016) and those expected from atmospheric scintillation noise alone (Osborn et al. 2015).

Figure 8 shows the median robust astrometric RMS as a function of airmass range, for the *g* and *r* filters using all fields observed from 2017-12-01UT to 2018-03-04UT. This includes data from the instrument commissioning period to sample data from the high-airmass experiments. Above airmass = 3, there is significant degradation in the astrometry. This degradation is due to a combination of the decrease in image quality (increase in FWHM) with increasing airmass causing source-centroid estimates to be less accurate in general and source-color dependent biases due to Differential Chromatic Refraction (DCR). Given ZTF's resolution, DCR starts to dominate above airmass $\simeq 1.8$. Currently, the primary survey is limited to airmass $\leqslant 2$ where the astrometric residuals are <65 and <85 milliarcsec RMS per-axis in *g* and *r* respectively. This is tolerable for most ZTF science programs. Airmass $\leqslant 2$ corresponds to altitudes of $\gtrsim 30°$ above the horizon and gives the requisite $3\pi$ sky coverage for the ZTF survey.

### 7.3. Photometric Precision (Repeatability)

Figure 9 quantifies the photometric repeatability as a function of magnitude by positionally matching sources from the quadrant-based PSF-fit catalogs extracted from $\sim 25$ to 95 image epochs. The overlapping images were observed at airmass $\lesssim 1.3$ and span 2018-02-04UT to 2018-03-04UT. Results are shown for four different quadrants selected from three different fields on the sky varying in source density. To mitigate variations in throughput (both atmospheric and instrumental), source fluxes in each catalog were globally rescaled to match those in the overlapping reference image (co-add) PSF-fit catalog. That is, each input epochal catalog was rescaled to a common zero-point equal to that of the reference image before the time-collapsed statistics were computed.

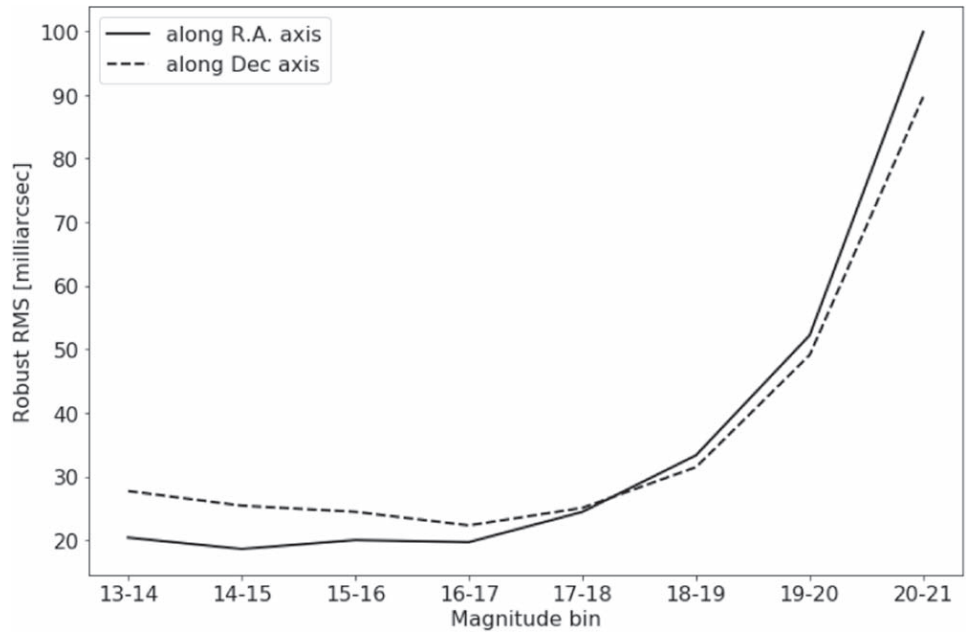

**Figure 7.** Robust astrometric RMS as a function of PSF-fit magnitude for a single CCD quadrant from a *g*-filter exposure. 350 stars fall in the 13–14 magnitude bin and 19,999 fall in the 20–21 magnitude bin. See Section 7.2 for details.

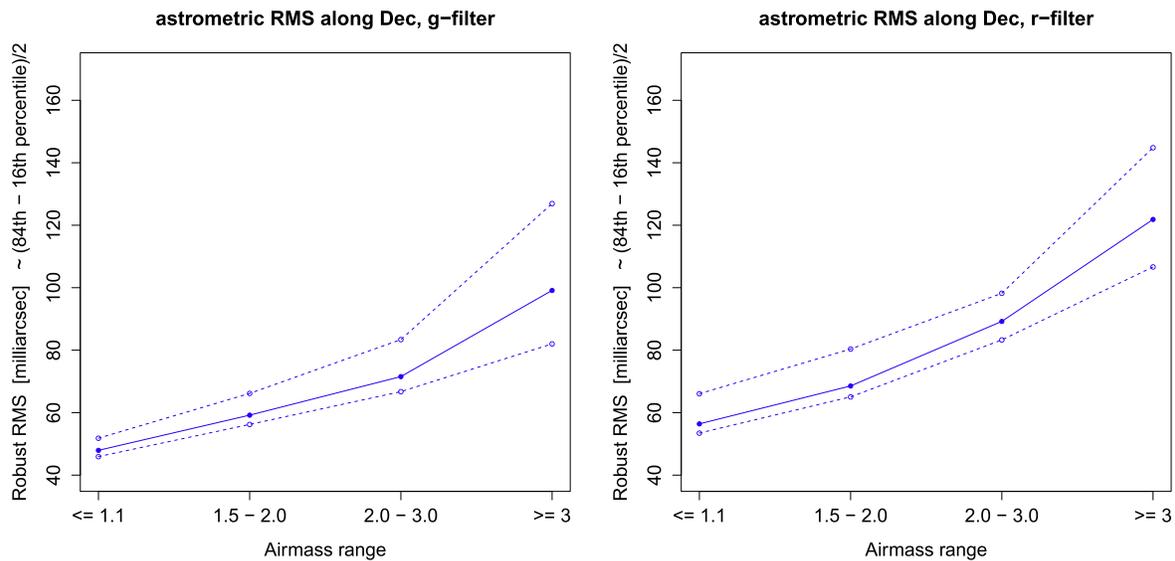

**Figure 8.** Astrometric RMS error in declination per CCD quadrant as a function of airmass range for *g*-filter *left* and *r*-filter *right*. The dependence in R.A. is similar. The middle filled circles (connected by solid lines) are median RMS values over all 64 CCD-quadrants; top open circles (with dashed lines) are maxima over all CCD-quadrants; bottom open circles (with dashed lines) are minima over all CCD-quadrants. These estimates pertain to ZTF extractions with photometric S/N $\geq$ 10. See Section 7.2 for details.
(A color version of this figure is available in the online journal.)

In general, the internal photometric precision derived from PSF-fit photometry at bright unsaturated fluxes typically lies between 8 and 25 millimag—the range encountered from examining about one hundred quadrant-based catalogs, independent of filter. The primary limiting factor that determines the resulting photometric precision for a CCD-quadrant is the accuracy of flat fields derived from the internal flat-screen exposures (Section 3.4.2). This was verified by constructing *low*-spatial frequency responsivity maps by binning the photometric residuals from source photometry and examining their spatial distribution.[55] These residuals are of approximately the correct magnitude to reduce systematics from inaccurate flat-fielding upstream and improve the photometric precision to consistently $\lesssim 10$ millimag at low to moderate airmass. Flat-fielding residuals have not yet been fully characterized for all CCD-quadrants and filters as a function of other factors (e.g., scattered light). There is a plan to augment the flat-screen *high*-spatial frequency responsivity maps and minimize the photometric residuals in general.

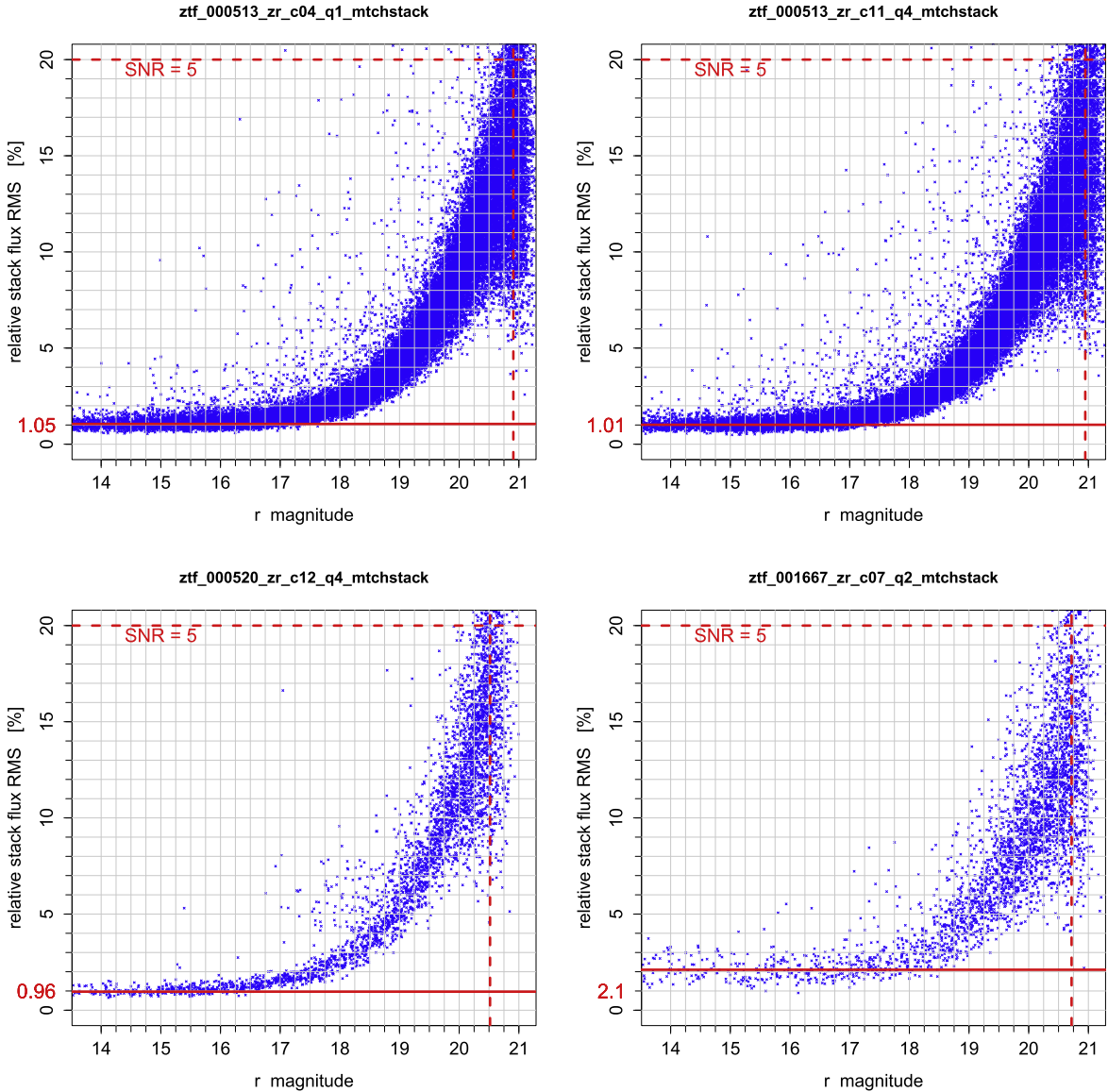

**Figure 9.** Relative flux RMS in stacked multi-epochal photometry (repeatability) using the PSF-fit catalogs from four CCD quadrants. The top two panels are for dense Galactic plane fields. Vertical dashed lines correspond to the approximate $5\sigma$ magnitude limits (20% flux RMS). Precision levels at bright fluxes are indicated on the y-axes. See Section 7.3 for details.
(A color version of this figure is available in the online journal.)

Cases with lower photometric precision in general ($\gtrsim 20$ millimag) mostly occur at higher airmass where image quality and astrometric accuracy are generally lower (Section 7.2). Furthermore, there is no significant deterioration in photometric precision in high-density source regions such as the Galactic plane (top two panels in Figure 9) where source de-blending is more prevalent.

---

[55] Both PSF-fit and aperture-photometry catalogs are used. The larger aperture measurements provide a valuable crosscheck. These are more immune to residual spatial variations in the PSF not caught by the PSF-estimation process prior to PSF-fitting.

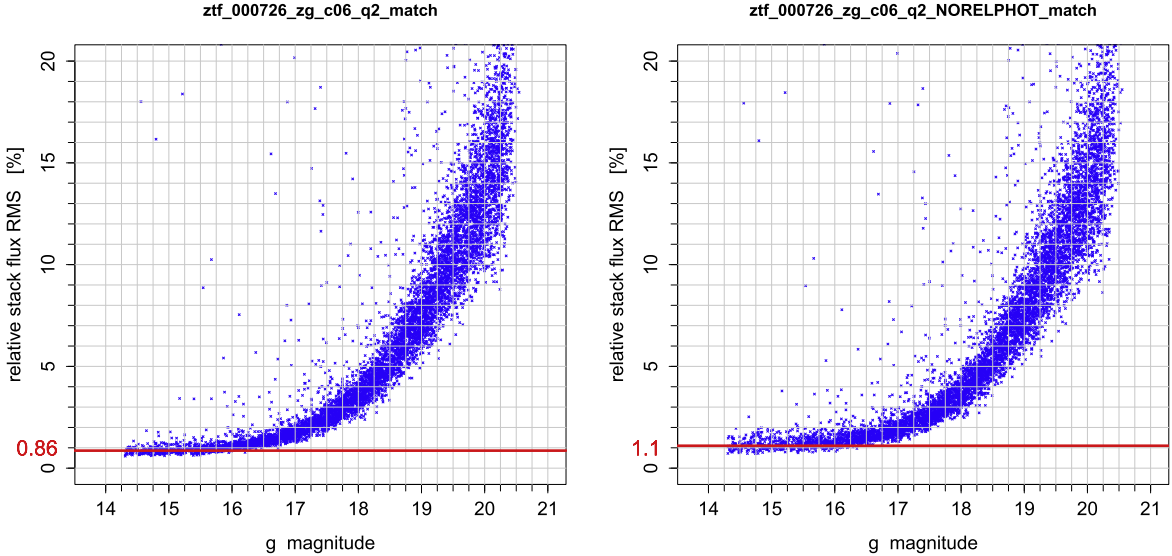

**Figure 10.** Relative flux RMS in stacked multi-epochal photometry (repeatability) from *matchfile* products with (left) and without (right) relative photometric refinement turned on. Precision levels at bright fluxes are indicated on the *y*-axes. See Section 7.3 for details.
(A color version of this figure is available in the online journal.)

Photometric repeatability metrics are readily available in the source *matchfile* products (Section 3.8), obviating the need for explicit matching of catalogs across epochs. These metrics can also be retrieved through the light-curve query service (Section 5). At the time of writing however, an insufficient number of *matchfiles* were available for sampling different source-density regions for use in quantifying the photometric precision (Figure 9). For comparison, example photometric repeatability plots from two *matchfiles* for the same CCD-quadrant are shown in Figure 10. Each was created in a slightly different way. The added benefit of *matchfile* creation is that it includes a post-processing step to mutually refine the individual photometric zeropoints in a relative sense (Section 3.8). Figure 10 shows the relative flux RMS with and without relative photometric refinement turned on. As a reminder, the results in Figure 9 (that matched raw catalog-sources directly) included no refinement of zeropoints. The improvement in relative precision from the refinement of photometric zero-points is typically 10% to 20%. Preliminary matchfile products show precision levels as low as $\sim$8 millimag per CCD-quadrant in either *g* or *r*. The achievable precision is limited by flat-fielding accuracy (see above) and other processing details that depend on sky location.

### 7.4. Sensitivity Limits

The photometric repeatability versus magnitude plots in Figure 9 can also be used to determine the $N\sigma$ magnitude limit for a specific survey field and CCD-quadrant. For instance, $5\sigma$ corresponds to a 20% flux RMS and from Figure 9, occurs at $r \simeq 20.5$ to 21.0 mag for the quadrants shown.

For comparison, Figure 11 shows the relative (model-based) flux uncertainty as a function of magnitude from two quadrant-based PSF-fit catalogs. This model uncertainty is based on propagating pixel uncertainties from a model involving the electronic gain, pixel RMS, PSF-estimation error, through to PSF-fitting. The $5\sigma$ (20% flux uncertainty) magnitude limits in Figure 11 are shown by the vertical intersecting lines. These are consistent with estimates from repeatability, implying the model photometric uncertainties are plausible. Furthermore, these estimates were validated in a global sense using matches to PS1 sources and their reported photometric errors.

The model-based flux-uncertainty versus magnitude from PSF-fit photometry (e.g., Figure 11) is used in general to estimate an approximate $5\sigma$ magnitude limit for each processed quadrant. This is based on computing the median magnitude of sources with S/N values falling within $4.5 \leqslant S/N \leqslant 5.5$. This magnitude limit is stored as the MAGLIM keyword in the FITS header of each processed CCD-quadrant image. Figure 12 shows the spatial variation in magnitude limit for each filter by medianing the values from several thousand quadrants in *g* and *r*, and a few hundred quadrants in *i* across 75 photometric nights spanning 2018-04-01UT to 2018-06-30UT, or $\simeq$3 lunations. All are at airmass $\leqslant$1.05 (i.e., near zenith) and all are from 30 sec exposures, the nominal integration time for ZTF.[56] The overall $5\sigma$ median magnitude limits across all quadrants and epochs used in these maps are $g = 21.067 \pm 0.003$, $r = 21.012 \pm 0.002$, and $i = 20.51 \pm 0.01$ mag. The $1\sigma$ uncertainties are based on $\sqrt{N}$ statistics.

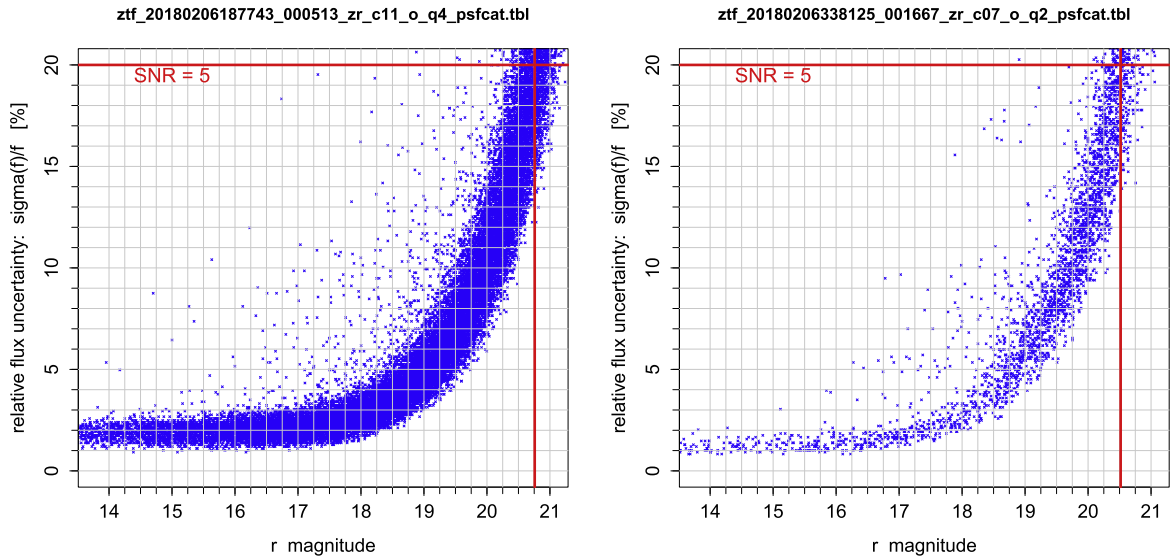

**Figure 11.** Relative (model-based) flux uncertainty as reported in two quadrant-based PSF-fit catalogs, for a dense Galactic plane field (left) and a high Galactic latitude field (right). Vertical lines correspond to the approximate $5\sigma$ magnitude limits (20% flux uncertainty cut). See Section 7.4 for details.
(A color version of this figure is available in the online journal.)

The spatial variations across CCD-quadrants in Figure 12 are consistent with variations in the electronic gain ($e-/$ADU) for each amplifier.[57] These are also consistent with variations in photometric throughput inferred from calibrated zeropoints. The drop in sensitivity in the corners is due to vignetting.

### 7.5. Photometric Calibration Assessment

The photometric calibration process was summarized in Section 3.5. Figure 13 shows the difference between PS1 and calibrated ZTF magnitudes for three quadrant-based PSF-fit catalogs in filters *g*, *r*, and *i*. The top panels show the residual magnitudes and the bottom panels are their corresponding RMS in flux space as a function of magnitude. The purpose here is to quantify the level of any photometric biases. Preliminary results show that there exist photometric biases in the PSF-fit catalogs of up to 0.02 mag for predominately bright sources. This represents the maximum median deviation measured at magnitude <15 in any filter from examining a few hundred PSF-fit catalogs (red circles in top panels of Figure 13). These biases are also field-dependent and analyses are underway to understand them. Similar biases are seen in the aperture-photometry-based catalogs;

---
[56] Some programs use integration times of 60 and 90 sec, and sometimes 300 sec.
[57] Inferred independently from analysis of illuminated flat-screen exposures.

however, these are generally larger in high source-density regions due to the effects of source crowding. PSF-fitting is more immune to this effect and hence, we recommend using the PSF-fit catalogs if one is interested in point-source photometry. The PSF-fit catalogs are more optimal for faint source photometry. For bright sources however, aperture photometry is generally superior, provided there is no contamination.

The bottom three panels in Figure 13 show the relative flux RMS-deviation with respect to PS1 in 0.5-magnitude bins. The scatter is dominated by the measured instrumental photometry, e.g., local background estimation, PSF estimation, flat-fielding, source confusion, and errors in global astrometry and centroiding. There is also some scatter in the PS1 measurements themselves. The RMS scatter with respect to PS1 is field-dependent and varies from $\sim 1\%$ to $\sim 2\%$ for bright, unsaturated sources. A global analysis of the overall photometric calibration accuracy as a function of all survey parameters is pending.

### 7.6. Difference Image Event Statistics

The extraction of point-source events from difference images was described in Section 3.6 and alert packets generated therefrom in Section 4.1. Figure 14 shows the number of *raw* (non-ML-vetted and non-human-vetted) point-source events to $5\sigma$ as a function of several image metrics at the CCD-quadrant level. These metrics refer to the instrumentally calibrated science image, whose photometric and astrometric solutions also apply to the derived subtraction images. Only quadrants containing at least one event from either their *positive* (science minus reference) or *negative* (reference minus science) subtraction image are plotted. The events were collected from eight continuous observing nights: 2018-05-23UT to 2018-05-30UT using exposures observed to an airmass of ≈1.5 in *g* and ≈2.5 in *r*. The observations include a mixture of integration times: 30 sec (nominal), 60 sec, and 90 sec. About 8% of the data are from fields observed at Galactic latitudes $|b| \lesssim 12°$.

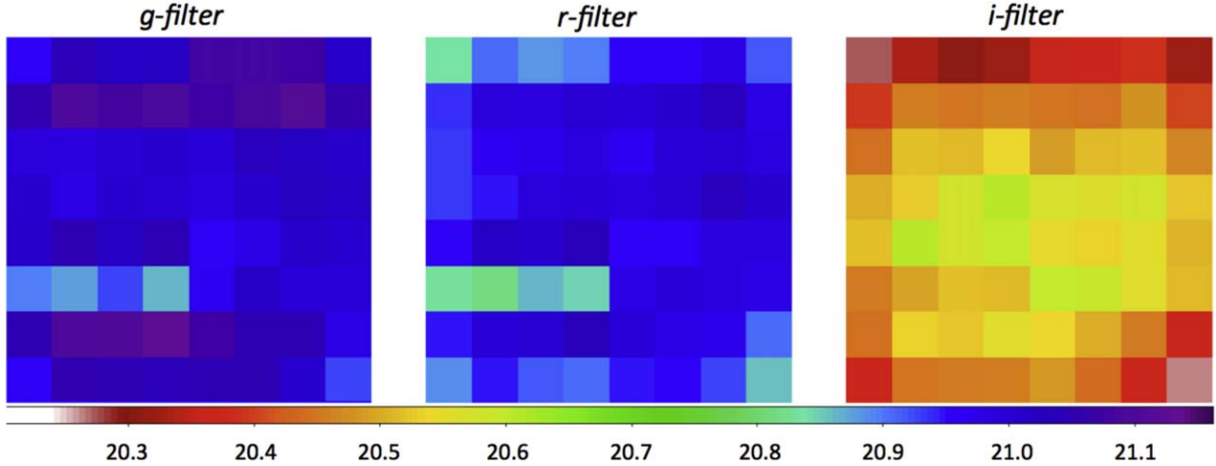

**Figure 12.** Focal plane maps of median MAGLIM for each CCD quadrant for *g, r, i* filters. Scale is shown on the color bar. North is up and east is to the right. See Section 7.4 for details.

Overall, the median number of *raw* point-source alerts per quadrant and per filter is ≈4 with a 95th percentile of ≈26 (*g*) and ≈43 (*r*). It is important to note that a significant fraction of these are false positives (from image artifacts; see Mahabal et al. 2018). We expect *at most* a few to several real astrophysical alerts per-image quadrant depending on sky location. For example, the alert density will be dominated by asteroids in the ecliptic plane and variable stars in the Galactic plane (see also the summary statistics in Section 6).

The clustering of points in the photometric-zero-point (ZP), median background, and to a lesser extent the magnitude-limit plot is due to different exposure times and observations in the Galactic plane. The ZP plot marks the median ZP values (as vertical lines) at the three exposure times: 30, 60, and 90 sec. The slope about each vertical line is due to varying airmass. Furthermore, values with ZP $\lesssim$ 25.7 mag in the ZP plot correspond to low atmospheric transparency (non-photometric conditions). The plot showing point-source FWHM (derived image quality) reflects the distribution of seeing and PSF sizes delivered by the P48 optics, convolved with $\simeq 1$ arcsec square pixels. A fraction of the images are under-sampled (FWHM < 2 arcsec). These images do not yield an excess of false events relative to over-sampled images. Such images are appropriately handled by our image subtraction and event-extraction algorithms (Section 3.6). The plot showing median radial separation from *Gaia* DR1 is based on PSF-fit extractions with photometric S/N $\geqslant$ 10 (see Section 7.2). This median radial separation is approximately equal to the RMS deviation per axis.

### 7.7. Latency of Alert Packet Generation

Figure 15 shows the typical timescales on which alert packets become available for querying at IPAC to support the ZADS (Section 4.1) following observation at the P48. Additional overheads from queries by downstream consumers, including community brokers amount to typically several seconds for ~1000 packets. This is further discussed in Patterson et al. (2019). Given that exposures are processed at the CCD-quadrant level (Section 2.1), the latency is tied to the processing of an individual quadrant image where difference-image-detected events are extracted and alert packets generated in one instance (single thread) of the real-time pipeline. This histogram was generated from a random sample of $\simeq$10,000 CCD-quadrants observed throughout June 2018. These yielded $\simeq$114,600 alert packets or $\simeq$11.5 alerts per CCD-quadrant on average.

The 5th, 50th, and 95th percentiles in the alert production latency are $\simeq$6.7, 8.5, and 12.6 minutes respectively. The tail is due to a combination of observations in ultra-dense regions of the galactic plane and additional overhead incurred from imperfect image differencing causing excessive numbers of events to be processed and filtered prior to packetizing. Approximately 25% of the exposures in this set were on dense regions of the galactic plane. The latency is dominated by pipeline processing ($\lesssim$5.8 minutes per CCD-quadrant; e.g., see Figure 5). The remaining time is due to data transfer from the P48, raw data ingestion and staging in the SLURM queue (Section 2.1), CCD splitting (Section 3.3), and further queueing prior to triggering of the real-time pipeline (Section 3.5, Section 3.6).

## 8. Advice and Lessons Learned

Below we list some of the challenges encountered during development of the ZSDS, leading into commissioning and survey operations. We consider the project to still be in its infancy in terms of interfacing with the wider community, although we have progressed far enough to highlight some key lessons. These reflect the authors' views and opinions, and not those of the project or its partners.

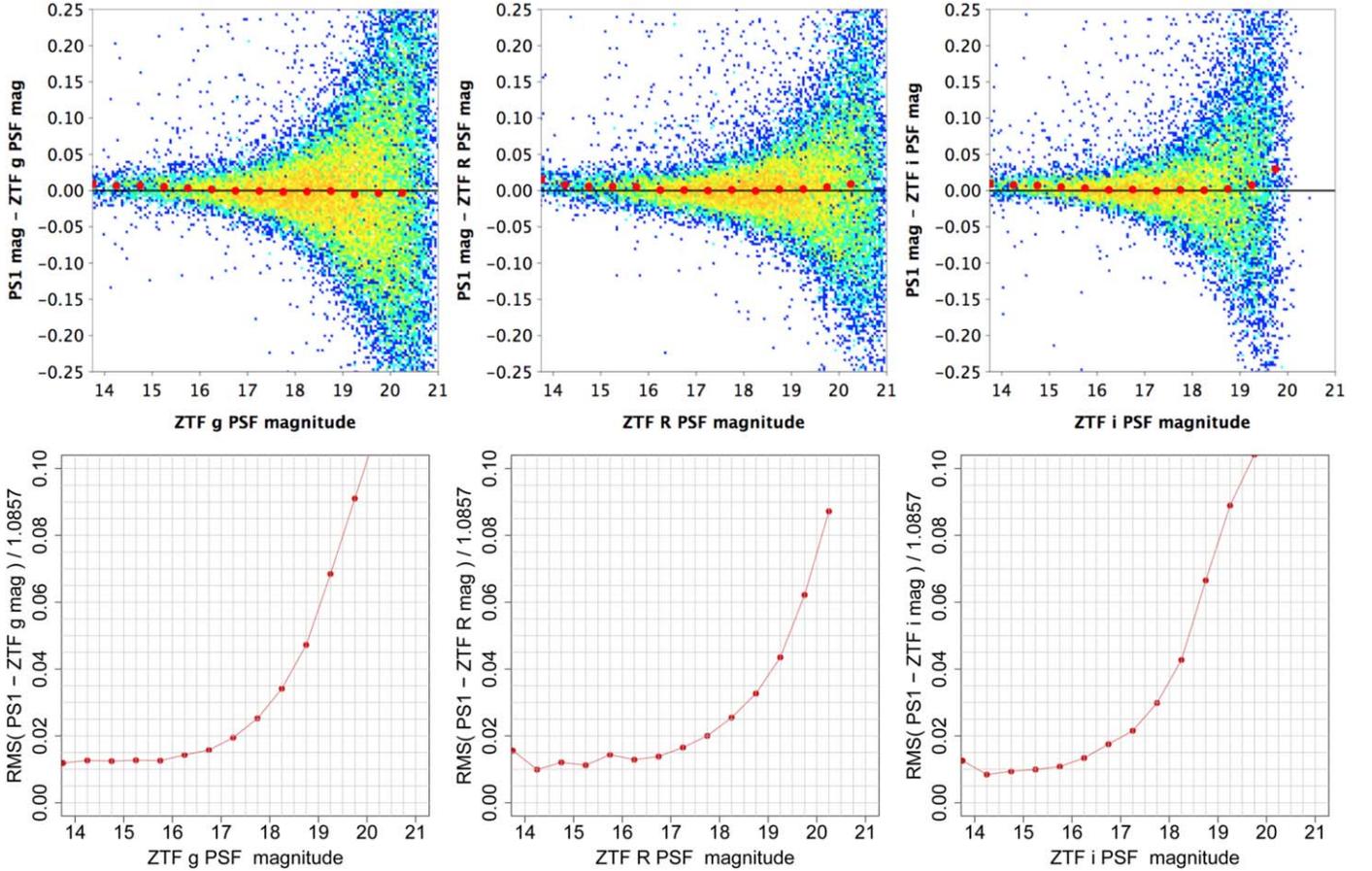

**Figure 13.** Top three panels: difference between PS1 and ZTF PSF-fit magnitudes for three epochal quadrant-based PSF-fit photometry catalogs for filters *g, r,* and *i* (left to right). The red circles represent binned, trimmed medians. The color range is from 1 (blue) to 32 (orange) sources per resolution element. Bottom three panels: corresponding relative flux RMS deviation using a *robust* measure with respect to PS1 ($=\sqrt{\langle (f_{ps1} - f_{ztf})^2 \rangle}/f_{ps1}$). The value 1.0857 ($\simeq 2.5/\ln[10]$) is used to convert from magnitude to flux space. The *g* and *r* plots (left and middle) correspond to Galactic plane observations.

1. ZTF is a big data, small-team project. Most of the deliverables and services are extensions of those from PTF and iPTF. One ongoing challenge is meeting the expectations of science users and justifying the reasons for our decisions. It is important to advertise all science requirements, product specifications, and capabilities according to available resources long before development begins. Engaging all project partners and the community early-on is crucial. There have been incidences of "creeping requirements" proposed during development and into commissioning for additional functionality and products. Once the foundation is in place, it is a costly exercise to re-architect the system on a short timescale to maintain overall processing efficiency. This required significant research and development and sometimes additional (unprovisioned) hardware.

2. Related to the previous point, it is important to have all interfaces between project subsystems documented and agreed-on in advance of any development. Of particular importance are interfaces of the ZSDS with the Observing System, QA/monitoring subsystem, and ZADS. Specifications and requirements need to be as detailed as possible. This includes all metrics and data product formats. All proposed updates need to be communicated, discussed, and evaluated against project resources before implementation.

3. Ensure tools are in place to assess the overall performance of image differencing and alert generation over the course of the survey. This includes tracking false-positive and false-negative rates. All necessary QA metrics as well as test survey fields for monitoring need to be identified. These test fields need to sample a range of astrophysical environments containing known reoccurring events (e.g., variable stars). The monitoring should include metrics to assess the quality of astrometric, photometric, and flat-field calibrations, all of which will impact the quality of difference images and events extracted therefrom.

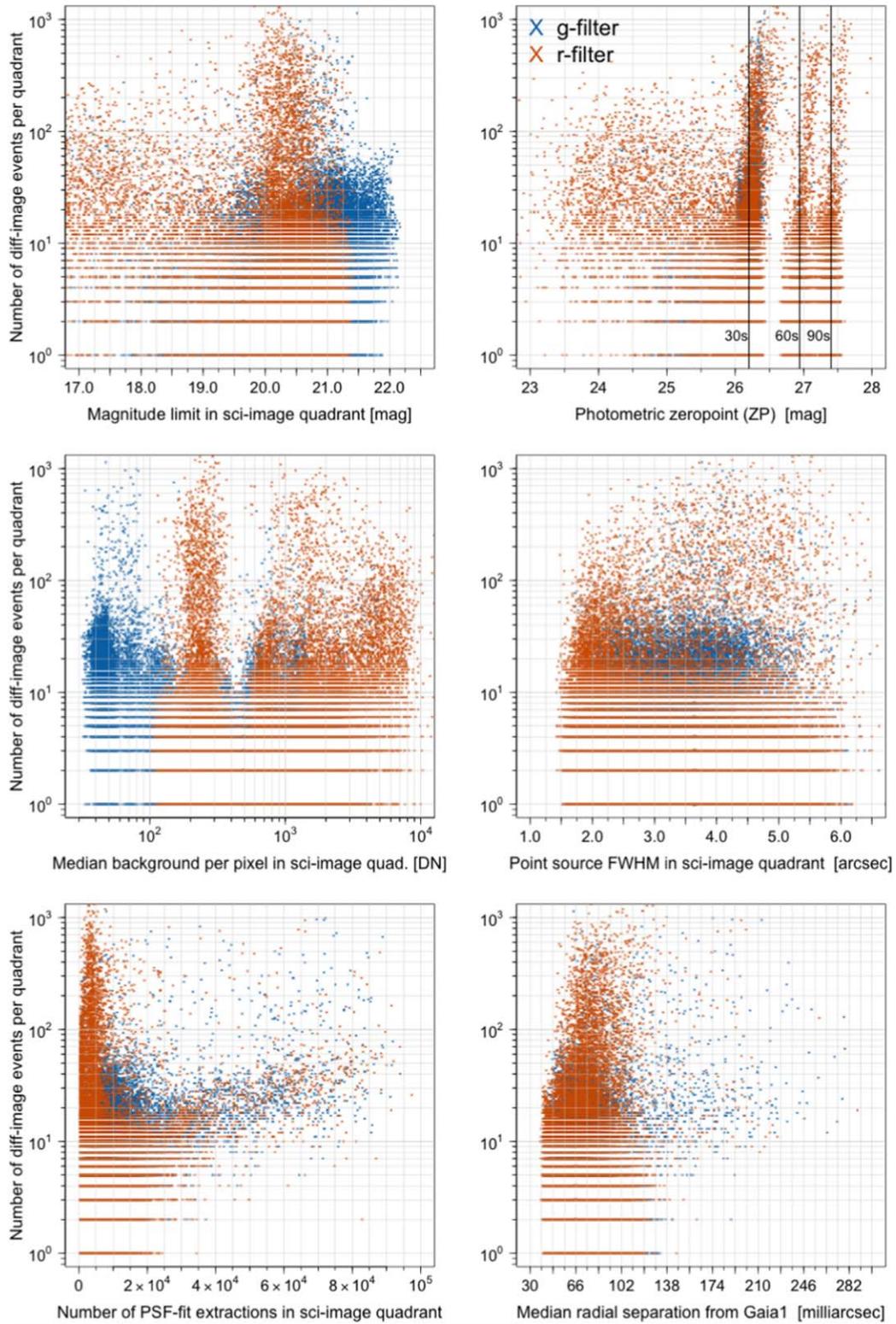

**Figure 14.** Number of difference image extracted events as a function of various science image metrics. Blue and vermillion (light brown) crosses represent *g* and *r* filter extractions respectively. For reference, a single science image quadrant covers ∼0.73 deg². See Section 7.6 for details.

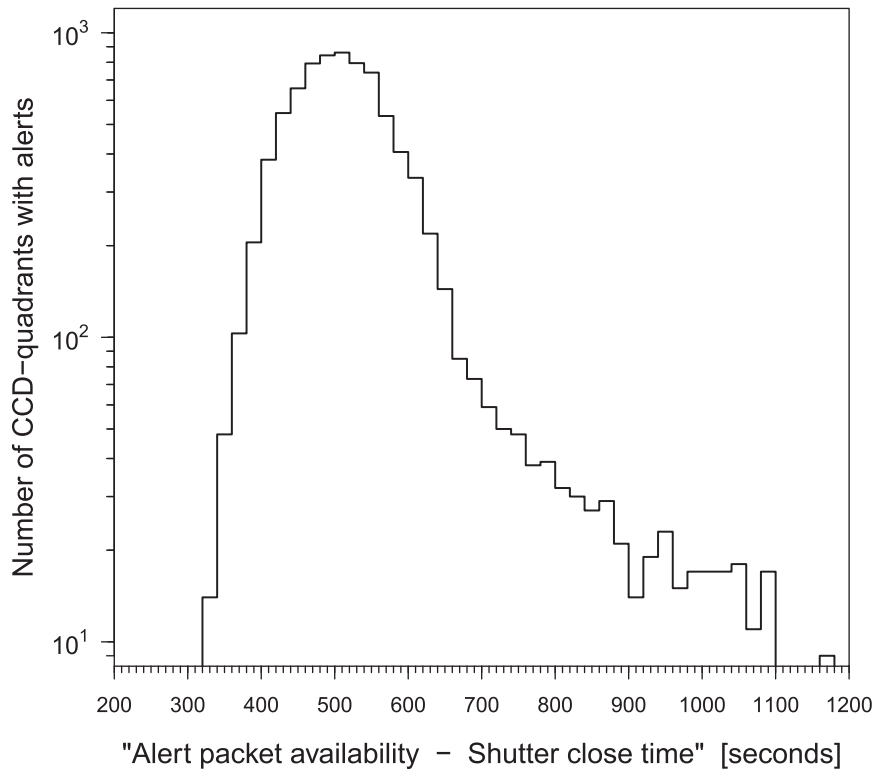

**Figure 15.** Distribution of the latency in the production of alert packets following observation. This includes transfer of raw data to IPAC, all processing steps, and staging of packets in IPAC's Kafka cluster, ready for querying by consumers. See Section 7.6 for details.

4. Related to the previous point, the reliability (and accuracy) of scores assigned to difference-image-detected events by ML-classifiers also needs to be monitored. In the context of *supervised* learning, the performance of a ML classifier depends crucially on the data on which it was trained. This should be re-evaluated due to unforeseen changes in instrumental signatures, observing conditions, and survey design. This includes updates to pipeline processing and reference images, and new (unencountered) astrophysical environments, i.e., not represented in the ML-training model.

5. We cannot overstate the importance of eyes-on-data during the commissioning and science-validation phases. One assumes that by simply focusing on "your" science, all will fall into place. We did not have enough eyes-on-data during these relatively short project phases. We recommend a systematic, coordinated effort to analyze all data products as a function of all observables encountered, including astrophysical environment. Getting more trained analysts involved early-on is a huge benefit.

6. Lastly, as a general comment, we advise hiring experienced engineers, scientific programmers, and data scientists whenever possible. In particular, those who strive for reliability and robustness. The ability to anticipate where things could go wrong and to implement safeguard measures is crucial in a fast-paced, big data project.

## 9. Summary and Future Updates

We have described the ZTF science data system, which encompasses the processing pipelines, real-time transient-alert distribution system, a discovery engine for SSOs, long-term archive, and the services to examine and retrieve the data products. Development was guided by our previous experience on PTF and is continuing to be optimized in response to the multitude of science programs currently underway.

Our design philosophy is flexibility and robustness. This includes being able to operate over a wide range of astrophysical environments and observing conditions, robustness to instrumental glitches, adaptability to changes in survey design, minimal tuning (unless warranted by changes in instrumental performance), resiliency to hardware outages, and efficient delivery of reliable astrophysical events for follow-up in near-real-time.

Ongoing improvements include refining ML algorithms for automatic vetting of point-source alerts and moving-objects, and refining instrumental calibrations to enhance astrometric and photometric quality as the survey proceeds. Additional functionality will include a service to perform forced-photometry on user-supplied target positions, and the ability to construct custom co-adds or mosaics from the image archive.

The ZSDS has benefitted from significant advancements in algorithms, software, and data-management practices at IPAC/Caltech. This paves the way to larger time-domain surveys, for example the Large Synoptic Survey Telescope, scheduled to begin operations in the early 2020s.

Based on observations obtained with the Samuel Oschin Telescope 48-inch and the 60-inch Telescope at the Palomar Observatory as part of the Zwicky Transient Facility project. Major funding has been provided by the U.S National Science Foundation under Grant No. AST-1440341 and by the ZTF partner institutions: the California Institute of Technology, the Oskar Klein Centre, the Weizmann Institute of Science, the University of Maryland, the University of Washington, Deutsches Elektronen-Synchrotron, the University of Wisconsin-Milwaukee, and the TANGO Program of the University System of Taiwan.

Part of this research was carried out at the Jet Propulsion Laboratory, California Institute of Technology, under a contract with the National Aeronautics and Space Administration.

The High Performance Wireless Research & Education Network (HPWREN; https://hpwren.ucsd.edu) is a project at the University of California, San Diego and the National Science Foundation (grant numbers 0087344 (in 2000), 0426879 (in 2004), and 0944131 (in 2009)).

This work has made use of data from the European Space Agency (ESA) mission *Gaia* (https://www.cosmos.esa.int/gaia),processed by the *Gaia* Data Processing and Analysis Consortium (DPAC, https://www.cosmos.esa.int/web/gaia/dpac/consortium). Funding for the DPAC has been provided by national institutions, in particular the institutions participating in the *Gaia* Multilateral Agreement.

This work has also made use of the Pan-STARRS1 (PS1) Surveys (http://pshttp://www.ifa.hawaii.edu/pswww/) and the PS1 public science archive (https://panstarrs.stsci.edu).

*Facilities:* PO:1.2 m, PO:1.5m.

# Appendix A
# Acronyms

| | | | |
|---|---|---|---|
| ADES | Astrometry Data Exchange Standard | NEA | Near-Earth Asteroid |
| ADU | Analog Digital Units | NED | NASA/IPAC Extragalactic Database |
| API | Application Programming Interface | NEOCP | Near-Earth Object Confirmation Page |
| ASCII | American Standard Code for Information Interchange | NFS | Network File System |
| CCD | Charge Coupled Device | OS | Observing System |
| CPU | Central Processing Unit | P48 | Palomar 48 inch |
| DAOPhot | Dominion Astrophysical Observatory Photometry package | PNG | Portable Network Graphics |
| DB | Database | PO | Palomar Observatory |
| DCR | Differential Chromatic Refraction | PS1 | Pan-STARRS Data Release 1 |
| $e-$ | Electrons | PSF | Point-spread Function |
| FTE | Full Time Equivalent | PTF | Palomar Transient Factory |
| FITS | Flexible Image Transport System | TPV | Tangent Parameter Value (distortion polynomial representation) |
| FOV | Field of View | RAM | Random Access Memory |
| FWHM | Full-width at Half-maximum | RMS | Root Mean Square deviation or error |
| GUI | Graphical User Interface | SDSC | San Diego Supercomputing Center |
| HDF5 | Hierarchical Data Formatted table file | SIMBAD | Set of Identifications, Measurements, and Bibliography for Astronomical Data |
| HDU | Header Data Unit | SLURM | Simple Linux Utility for Resource Management |
| IO | Input/Output | SNR | Signal-to-Noise Ratio (sometimes abbreviated as S/N) |
| IPAC | Infrared Processing and Analysis Center | SSO | Solar-system Object(s) |
| IRSA | NASA/IPAC Infrared Science Archive | SQL | Structured Query Language |
| JPEG | Joint Photographic Experts Group | QA | Quality Assurance |
| LUT | Look Up Table | URL | Uniform Resource Locator |
| MBA | Main Belt Asteroid | UW | University of Washington |
| MEF | Multi-Extension FITS file format | VO | Virtual Observatory |
| ML | Machine-learning (or Machine-Learned) | VPO | Virtual Pipeline Operator |
| MOPS | Moving Object Processing System | WCS | World Coordinate System |
| *MOST* | Moving Object Search Tool | XML | Extensible Markup Language |
| MOV | Movie file format | ZADS | ZTF Alert Distribution System |
| MPC | Minor Planet Center | ZMODE | ZTF Moving Object Discovery Engine |
| MSIP | Mid-scale Innovations Program | ZOGY | Zackay Ofek Gal-Yam (image subtraction algorithm) |
| NAIF | NASA's Navigation and Ancillary Information Facility | ZP | ZeroPoint (photometric zero-point) |
| NASA | National Aeronautics and Space Administration | ZSDS | ZTF Science Data System |

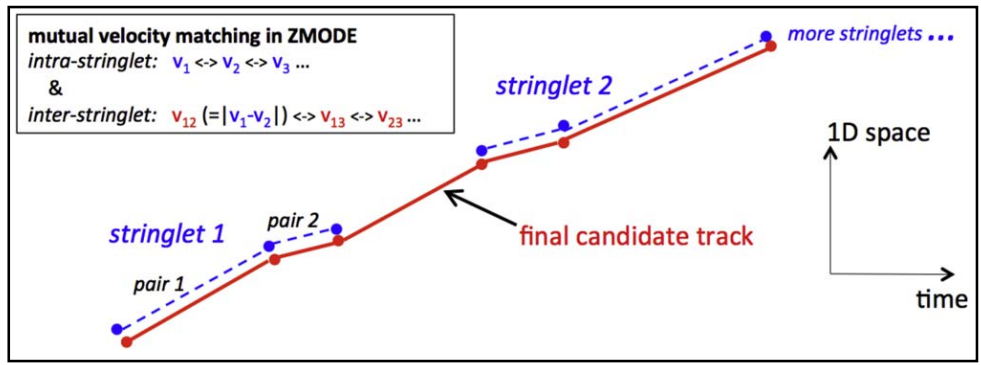

**Figure 16.** Schematic showing the construction of moving-object tracks in ZMODE.

# Appendix B
# ZMODE: ZTF's Moving Object Discovery Engine

ZMODE was developed at IPAC for use on the Palomar Transient Factory (PTF). It was developed for use on detections from difference images, with provisions for scaling up to meet ZTF's computational demands by leveraging new hardware architectures. It uses a new algorithm, conceived from ideas presented in Waszczak et al. (2013). Studies of the recovery fraction of known SSOs in PTF data show that ZMODE can detect objects with an efficiency of >90% and a reliability of >98% to moderately faint flux levels ($R_{PTF} \sim 20$ mag).

The most significant difference between ZMODE and the classic Moving Object Processing System (MOPS; Denneau et al. 2013) implementation is the first step in linking candidate moving-object detections into building blocks of candidate tracks. ZMODE requires a minimum of three detections to form a moving-object "stringlet", whereas MOPS starts out by forming two-detection "tuples". ZMODE forms stringlets by matching the relative velocities of two adjacent pairs of detections that share a common middle detection. A schematic is shown in Figure 16. The velocities are matched within some tolerance that depends on the time separation of the detections. The stringlet-construction step imposes two constraints on the time separation of detections, i.e., their observation epochs: (i) one of the detection pairs in the stringlet must span ⩽10 hours (i.e., the same night); and (ii) the other pair must span ⩽2.5 days. Criterion (ii) is optimal for input data spanning four consecutive nights. Four consecutive nights is driven by ZTF's observing cadence in the *public* portion of the survey and a minimum track length (number of detections) of four to declare a candidate track for vetting (see below). Detections across all ZTF filters (*g*, *r*, *i*) are linked when generating tracks.

After all possible stringlets in the detection stream have been identified, they are linked using velocity matching on a coarser grid to create moving-object candidate tracks. Two types of velocities are used in the final stringlet-merging step: (i) the mean velocity of the (intra-stringlet) detection pairs; and (ii) the relative (inter-stringlet) velocities between the average position of detections in each of the stringlets. The candidate tracks from ZMODE are then vetted using a number of quality metrics, including an orbit fit to test if the candidate track is dynamically plausible (see below).

By building three-detection stringlets, ZMODE moves part of the combinatorial challenge earlier in the processing. It helps eliminate spurious two-detection "tuples" that would otherwise add to the load at later stages of candidate track construction. Moreover, it significantly reduces the number of possible combinations that need to be processed and merged downstream, as opposed to MOPS which carries along all possible two-detection "tuples". The ZMODE design is therefore more efficient, for example, in regions of high source density for a given input timespan.

To reduce the incidence of false linkages, we automatically identify reliable tracks using orbit fitting. We find that tracks consisting of four or more linked detections accurately fitting a dynamically valid orbit constitute worthy candidates for further inspection. Orbit fitting is based on the find_orb[37] software, which has been adapted and tuned to process candidate tracks from ZMODE. Find_orb generates an orbit quality metric (or score) along with an RMS of the fit residuals. These two metrics are thresholded to retain good-quality tracks for both MBAs and NEAs.

The ZMODE algorithm and its performance will be described in more detail in a future publication.

## ORCID iDs

Frank J. Masci  https://orcid.org/0000-0002-8532-9395